\documentclass[onecolumn]{IEEEtran}
\usepackage{graphicx}
\usepackage{amsmath,amssymb,dsfont,bbm} 
\usepackage{color}
\usepackage{cite}
\usepackage{graphics}
\graphicspath{{.}{./Figures/}} 
\allowdisplaybreaks[4] 

\newtheorem{lemma}{Lemma}
\newtheorem{corollary}{Corollary}
\newtheorem{theorem}{Theorem}
\newtheorem{proposition}{Proposition}
\newtheorem{remark}{Remark}
\newtheorem{definition}{Definition}
\newtheorem{example}{Example}

\newcommand{\nintegers}{{\mathbb{Z}^+}}
\newcommand{\pintegers}{{\mathbb{Z}^+}}

\newcommand{\N}{{\mathbb Z^+}}
    
\newcommand{\Z}{{\mathbb Z}}         
  
\newcommand{\R}{{\mathbb R}}        
\newcommand{\bs}[1]{^{\hspace{-0.2mm}\scriptscriptstyle (\hspace{-0.2mm}#1\hspace{-0.2mm})}}
\newcommand{\h}[1]{{\hat{#1}}}

\renewcommand{\Pr}{{\mathbb{P}}}
\newcommand{\s}[1]{{\mathcal{#1}}} 
\newcommand{\sK}{{\{1,\ldots,K\}}} 
\newcommand{\sRk}{{\{1,\ldots,\lfloor 2^{n R_k} \rfloor \}}}
\newcommand{\RTx}[3]{{R\bs{#1}_{\text{Tx}, #2 \to #3}}}
\newcommand{\sRTx}[3]{{\Big\{1,\ldots,\big\lfloor 2^{n \RTx{#1}{#2}{#3}} \Big\rfloor \Big\}}}
\newcommand{\RRx}[3]{{R\bs{#1}_{\text{Rx}, #2 \to #3}}}
\newcommand{\sRRx}[3]{{\Big\{1,\ldots, \Big\lfloor 2^{n \RRx{#1}{#2}{#3}} \Big\rfloor \Big\}}}

\newcommand{\muT}{{\mu}_\Tx}
\newcommand{\muR}{{\mu}_\Rx}

\newcommand{\kappaT}{{\kappa_\text{Tx}}}
\newcommand{\kappaR}{{\kappa_\text{Rx}}}

\newcommand{\piT}{{\pi}_\Tx}
\newcommand{\piR}{{\pi}_\Rx}

\newcommand{\bfkappa}{{\boldsymbol{\kappa}}}
\newcommand{\bfmu}{{\boldsymbol{\mu}}}
\newcommand{\bfR}{{\boldsymbol{R}}}

\renewcommand{\S}{{\sf{S}}}

\newcommand{\Sach}{{\sf{S}_{\sf{Ach}}}}

\newcommand{\Ca}{\sf{C}}
\newcommand{\Cs}{\sf{C}_\Sigma}

\renewcommand{\i}{{\iota}}

\newcommand{\all}{\textnormal{all}}
\newcommand{\even}{\textnormal{even}}
\newcommand{\odd}{\textnormal{odd}}
\newcommand{\noises}{\textnormal{noises}}
\newcommand{\eqdef}{\triangleq}

\newcommand{\Tx}{\textnormal{Tx}}
\newcommand{\Rx}{\textnormal{Rx}}
\newcommand{\RT}{{R_\Tx}}
\newcommand{\RR}{{R_\Rx}}

\newcommand{\RRj}[1]{{R^{(#1)}_{\text{Rx}, k\to k'}}}

\newcommand{\tL}{{t_\text{L}}}
\newcommand{\tR}{{t_\text{R}}}

\definecolor{ForestGreen}{rgb}{0.0, 0.5, 0.0}

\begin{document}
\title{\huge Conferencing in Wyner's Asymmetric Interference Network: Effect of Number of Rounds}
\author{Mich\`ele~Wigger, Roy Timo and Shlomo Shamai (Shitz)
\thanks{The results in this paper were partly presented at the {2015 IEEE Information Theory Workshop, Jerusalem, Israel.} \newline
M.~Wigger is with the Signal and Information Processing Laboratory at Telecom ParisTech, michele.wigger@telecom-paristech.fr. R.~Timo is with the Institute for Communications Engineering at the Technische Universit\"{a}t M\"{u}nchen, roy.timo@tum.de. S.~Shamai is with the Department of Electrical Engineering, Technion Israel Institute of Technology, sshlomo@ee.technion.ac.il. \newline
S. Shamai was supported by the Israel Science Foundation (ISF) and the FP7 Network of Excellence in Wireless COMmunications NEWCOM\#. M. Wigger was supported by the city of Paris under the program ``Emergences.'' R.~Timo was supported by the Alexander von Humboldt Foundation. Some of the material in this paper was presented at the IEEE Information Theory Workshop (ITW), Jerusalem, Israel, April, 2015.}
}
\maketitle


\begin{abstract}
Our goal is to study the effect of the number of conferencing rounds on the capacity of large interference networks. 
We do this at hand of the per-user multiplexing gain (MG) of  Wyner's soft-handoff model with dedicated conferencing links between neighbouring transmitters and  receivers. We present upper and lower bounds on the per-user MG of this network, which depend on the capacities of the transmitter- and receiver- conferencing links and on the number of allowed conferencing rounds. The bounds are tight when: the prelogs of the conferencing links are small or  high; there is only transmitter conferencing or only receiver conferencing; or some symmetry conditions between transmitter-conferencing and receiver-conferencing hold. We also determine the per-user MG of the network when the number of conferencing rounds is unlimited. 

Our results show that for small conferencing prelogs $\lesssim 1/6$, a single conferencing round suffices to attain the maximum per-user MG when the number of conferencing  rounds is unconstrained. In contrast, when the prelogs are large, then every additional conferencing round increases the maximum per-user MG. 
\end{abstract}


\section{Introduction}

We study a communication network where transmitters and receivers can cooperate over dedicated links  that do not interfere with the main communication over the network. 
We analyse how the performance over a communication network depends on the number of interactive communication rounds that are allowed to take place over  the conferencing (cooperation) links. 
 Protocols with only few {conferencing rounds} are advantageous in practice because they can also be implemented when the conferencing communication is subject to stringent latency constraints or when the cooperative devices are limited in storage and computational capabilities.

Willems seminal work \cite{willems83}  shows that  a single conferencing round---during which the transmitters exchange parts of their messages---is optimal  when the network is a  two-user discrete memoryless multi-access channel (MAC). The same applies also to the two-user memoryless Gaussian MAC \cite{brosslapidothwigger12}; the three-user Gaussian or discrete memoryless MAC where the transmitters cooperate over ``public links" that are observed by all transmitters \cite{kramerwigger09}; and for the compound discrete memoryless MAC \cite{maricyateskramer07}. For the three-user memoryless Gaussian MAC with ``private" conferencing links where each transmitter can send cooperation information only to its left-neighbour, two conferencing rounds---during which the transmitters share and relay parts of their messages---are sum-rate optimal at high signal-to-noise ratio (SNR) \cite{simeoneetal08}.

A single conferencing round is also optimal for the physically degraded two-user  discrete memoryless broadcast channel (BC) with conferencing receivers \cite{daboraservetto06-2}. For general broadcast channels this does not seem to be the case, and Dabora and Servetto proposed an interactive two-round conferencing protocol \cite{daboraservetto06-2}. Interactive two-round conferencing protocols have also been proposed for two-user discrete memoryless BCs for scenarios where both receivers are interested in the same message or where only one of the two receivers has a message to decode \cite{ng}.


Conferencing has been studied in many other scenarios, e.g., \cite{mao, sadaf, ziv, yossi, ihsiang-tx, ihsiang-rx,ng} generally restricting attention to small networks and to one or two conferencing rounds. 

In this work we consider a large interference network with an arbitrary  large number $K$ of transmitters and receivers and we analyze on how the performance over the network  depends on the number of allowed conferencing rounds between transmitters and between receivers.

Specifically, we consider Wyner's asymmetric \emph{soft-handoff} model \cite{Hanly-Whiting-93,Wyner-94,Simeone-2011-Tut}  in Figure~\ref{Fig:AsyWynerNet} where the $K$ transmitters and $K$ receivers are aligned on opposite grids and each transmitted signal is received at its corresponding receiver and the receiver to its right. Each transmitter and each receiver can cooperate with its immediate left and right-neighbours over non-interfering dedicated conferencing links of capacities $\muT\cdot \frac{1}{2}\log (1+\textnormal{SNR})$ and $\muR \cdot\frac{1}{2} \log(1+\textnormal{SNR})$, where SNR stands for signal-to-noise ratio. Cooperation between transmitters can take place over at most $\kappa_\Tx$ rounds and conferencing between receivers over at most $\kappa_\Rx$ rounds.  For the setup in Figure~\ref{Fig:AsyWynerNet} this makes that information about a transmitter's message can propagate only to its $\kappaT$ left- and right-neighbours, and information about a receiver's output signal can propagate only to its $\kappaR$ left- and right-neighbours. 

Our measure of performance is the high-SNR sum-capacity of the Wyner network, more specifically,  the asymptotic per-user multiplexing gain (MG). 
We propose upper and lower bounds on the per-user MG (Theorems~\ref{Thm:Converse} and \ref{Thm:Ach}), which depend on the conferencing prelogs $\muT$ and $\muR$ and on the number of allowed conferencing rounds $\kappaT$ and $\kappaR$. Our bounds coincide when: 
\begin{itemize} 
\item conferencing prelogs $\muT$ and $\muR$ are below given thresholds that depend on the number of allowed conferencing rounds $\kappaT$ and $\kappaR$ (Corollary~\ref{Cor:SmallMu});
\item conferencing prelogs $\muT$ and $\muR$ exceed given thresholds that again depend on the number of allowed conferencing rounds $\kappaT$ and $\kappaR$ (Corollary~\ref{Cor:LargeMu});
\item only transmitters can cooperate or only receivers can cooperate, i.e., $\muR=0$ or $\muT=0$ (Corollary~\ref{cor:e}); and
\item the symmetry condition $\frac{\muT}{\kappaT}=\frac{\muR}{\kappaR}$ between transmitter and receiver conferencing holds (Corollary~\ref{Cor:Sym}).
\end{itemize} 

For comparison, we also derive the per-user MG when the number of conferencing rounds is not limited (Theorem~\ref{Thm:UnlimitedConf}).

Our results show that for small  conferencing prelogs, in particular for $\muT,\muR \leq 1/6$, a single conferencing round at transmitters and receivers allows to achieve the same  per-user MG as when the number of conferencing rounds is unconstrained. In contrast, for large conferencing prelogs, the maximum per-user MG increases with every additional conferencing round. In particular, with $\kappaT$ transmit and $\kappaR$ receive conferencing rounds, the per-user MG saturates at $\frac{2\kappaT+2\kappaR+1}{2\kappaT+2\kappaR+2}$ for large values of $\muT$ and $\muR$.

To the best of our knowledge, these results are the first to quantify the impact of the number of conferencing rounds on the capacity of a network.

In interference networks, both transmitter cooperation and receiver cooperation allow to mitigate interference \cite{ihsiang-tx, ihsiang-rx, ntranosmaddah-alicaire14-1, maricyateskramer07, lapidothshamaiwigger07, levyshamai09, lapidothlevyshamaiwigger14,biao, bandeelgamalveeravalli15}. Recently, Ntranos, Maddah-Ali, and Caire~\cite{ntranosmaddah-alicaire14-1} proposed the following cooperation protocol for interference mitigation in large Gaussian networks: transmitters share quantised versions of their transmit signals and receivers share parts of their decoded messages.  (This is different from many previous works where transmitters share parts of their messages and receivers share quantised versions of their received signals.) Knowledge about other transmitters' input signals allows the transmitters to mitigate the interference caused by these signals using dirty-paper coding. Similarly, knowledge of decoded messages allows receivers to reconstruct interferences and subtract them from received signals. 

A drawback of these interference mitigation techniques is the delay in communication and the propagation of interference they induce. Specifically, each transmitter~$k$ has to wait until it has obtained the quantisation information about the transmit signals it wishes to mitigate,  before it can construct its own input signal $X_k^n$ and send quantisation information about it to  its other neighbours. 
In a similar way, each receiver has to wait until it obtains the decoded messages pertaining to some of the transmitters that interfere its output signals, before it can decode its own message and send parts of it to its other neighbours. 

%

Our coding scheme presented in this paper is inspired by the Ntranos, Maddah-Ali, and Caire  \cite{ntranosmaddah-alicaire14-1} protocol explained above. (Transmitters share quantised versions of transmit signals and receivers share parts of decoded messages.) The constraints on the number of cooperation rounds $\kappaT$ and $\kappaR$ however require the following two major changes: 
\begin{itemize} 
\item Since the transmitters and receivers cannot wait infinitely long before producing their cooperation messages, the protocol can only be applied over subsets of the network. 
 This necessitates to periodically silence transmitters in the network, which decomposes the network into smaller subnets. 
\item The interest is to switch off as few transmitters as possible and thus to make the subnets as large as possible. To this end, a sophisticated combination of the described transmitter and receiver interference-mitigation techniques is required. We also introduce slight variations of these techniques that can exploit the cooperation links from the transmitters to their left neighbours and from the receivers to their left neighbours.
\end{itemize}

For the case $\kappaT=\kappaR=1$, the performance of our schemes can also be achieved using conferencing protocols where the transmitters share parts of their messages and receivers share quantisation information about their receive signals. The advantage of these protocols is that they do not necessitate codebook knowledge during the conferencing phase and can thus be implemented in \emph{oblivious scenarios}. 

\subsection{Notation}

We denote the integers by $\Z$, the positive integers by  $\N$, and the real numbers by $\R$. Random variables are identified by uppercase letters, e.g. $W$, their alphabets by matching calligraphic font, e.g. $\s{W}$, and elements of an alphabet by lowercase letters, e.g. $w \in \s{W}$. The Cartesian product of $\s{W}$ and $\s{W}'$ is $\s{W} \times \s{W}'$, and the $n$-fold Cartesian product of $\s{W}$ is $\s{W}^n$. For any $n$-tuple of random variables $W_1,\ldots, W_n$ we use the shorthand notation $W^n:=(W_1,\ldots, W_n)$.

 Given two $n$-dimensional vectors ${a}^n,{b}^n\in\R^n$, let $\|a^n\|$ denote the standard norm of $a^n$ in Euclidean space,  and let $<a^n, b^n>$ denote the standard inner product of $a^n$ and $b^n$. Let further $\angle({a}^n,{b}^n)$ denote the angle between the two vectors:
 \begin{equation*}
 \angle(a^n,b^n):=\arccos \frac{<a^n,b^n>}{\|a^n\| \|b^n\|},
 \end{equation*}
 where $\arccos$ denotes the arc-cosine function.

\subsection{Organisation of Paper}

The rest of this paper is organised as follows. Section~\ref{sec:DescriptionOfTheProblem} describes the problem setup. Section~\ref{Sec:MainResults} states the main results of the paper. Proofs  of the results are presented in Sections~\ref{sec:converse_unlimited}--\ref{Sec:Proof:Thm:Ach}: Section~\ref{sec:converse_unlimited} proves the achievability of our Theorem~\ref{Thm:Ach}; Section~\ref{Sec:Proof:Thm:Converse} proves our converse Theorem~\ref{Thm:Converse}; and Section~\ref{Sec:Proof:Thm:Ach} proves the converse to Theorem~\ref{Thm:UnlimitedConf}.


\section{Problem Setup}\label{sec:DescriptionOfTheProblem}


\subsection{Channel Model and Transmit Power Constraint}

\begin{figure}
\begin{center}
\includegraphics[width=0.6\textwidth]{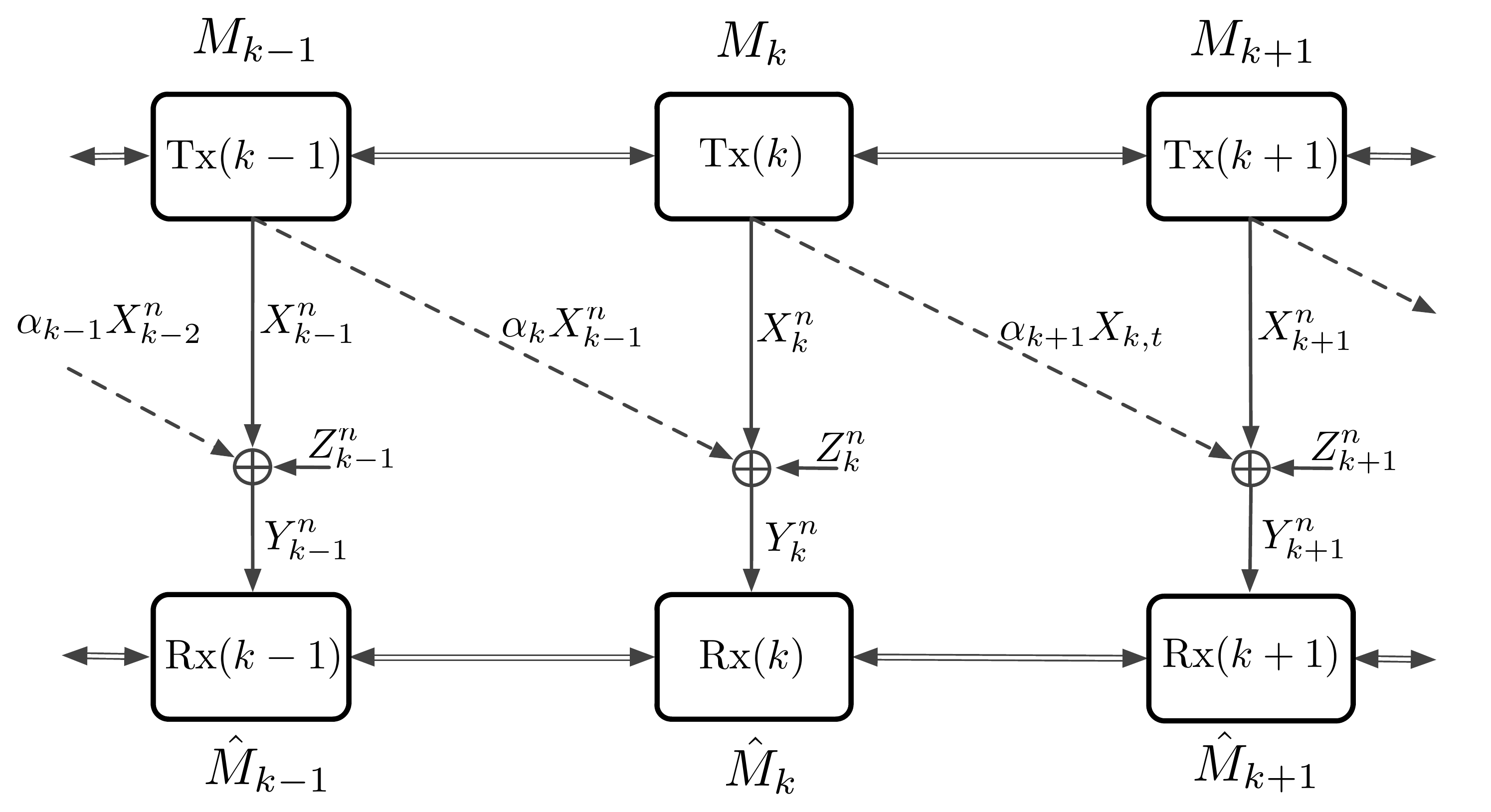}
\caption{Wyner's asymmetric interference network with rate-limited noiseless cooperation links between neighbouring transmitters and neighbouring receivers.}
\label{Fig:AsyWynerNet}  
\end{center}
\end{figure}

Consider a wireless communications system with $K$ pairs of transmitters and receivers, labeled by $k \in \sK$. Assume that the transmitters and receivers are each equipped with a single antenna, and that all channel inputs and outputs are real valued. We imagine a network with short-range interference, \emph{\`a la~}\cite{Wyner-94, Hanly-Whiting-93, rate-limited, lapidothlevyshamaiwigger14}, so that the signal sent by transmitter~$k$ is only observed by receivers~$k$ and $k+1$. Specifically, the time-$t$ channel output at receiver~$k$ is 
\begin{equation}\label{Eqn:Channel}
Y_{k,t} = X_{k,t} + \alpha_k X_{k-1,t} +Z_{k,t},
\end{equation}
where $X_{k,t}$ and $X_{k-1,t}$ are the symbols sent by transmitters~$k$ and $k-1$ at time $t$ respectively; $\{Z_{k,t}\}$ are independent and identically distributed (i.i.d.) standard Gaussians for all $k$ and $t$; $\alpha_k\neq 0$ is a fixed real number; and $X_{0,t} = 0$ for all $t$. A small segment of this  short-range interference model is depicted in Figure~\ref{Fig:AsyWynerNet}.

Each and every transmitter~$k \in \sK $ is required to reliably communicate a source message $M_k$ to its corresponding receiver~$k$. The source message $M_{k}$ is uniformly distributed on
\begin{equation*}
\mathcal{M}_{k} := \sRk
\end{equation*}
where $n$ denotes the blocklength and $R_k$ the rate of transmission of source message $M_k$.
All source messages are  independent of each other and of all channel noises. An average block-power constraint $P>0$ is imposed on the transmitted signals: 
\begin{equation}\label{eq:power}
\frac{1}{n} \sum_{t=1}^n X_{k,t}^2
\leq P, \quad \text{a.s.},
\quad \forall\ k \in \sK.
\end{equation}


\subsection{Overview of Conferencing and Communications Phases}

A key feature of this work is that we include \emph{rate-limited local cooperation} between neighbouring transmitters and neighbouring receivers over dedicated conferencing links. Specifically, we suppose that the communications process consists of the following four phases.
\begin{enumerate}

\item \emph{Tx-Conferencing Phase:} The source messages $(M_1\ M_2\ \ldots\ M_K)$ are revealed to their respective transmitters, and each transmitter exchanges conferencing messages with its two direct neighbours over dedicated noiseless channels. Each of these conferencing channels has a maximum rate budget of $n\RT$ bits. We let the rate budget $\RT$ scale with the transmit power constraint $P$ as
\begin{equation}\label{Eqn:TxRateBudget}
\RT :=  \muT\ \frac{1}{2}\log(1+P),
\end{equation}
where $\muT \in [0,\infty)$ is fixed and called the \emph{Tx-prelog} conferencing constant. 

\item \emph{Cooperative Communications Phase:} The transmitters communicate their source messages over the memoryless interference channel~\eqref{Eqn:Channel}.  Transmitter~$k$'s channel inputs are a function of its source message $M_k$ and the conferencing messages it received during the Tx-conferencing phase. 

\item \emph{Rx-Conferencing Phase:} The receivers observe their channel outputs, and they exchange conferencing messages with their immediate neighbours over dedicated noiseless channels. Each of these conferencing channels has a maximum rate 
\begin{equation}\label{eq:RRx}
\RR := \muR\ \frac{1}{2} \log (1 + P).
\end{equation} 
Here $\muR \in [0,\infty)$ is fixed and called the \emph{Rx-prelog} conferencing constant. 

\item \emph{Decoding Phase:} The receivers decode their desired source messages from the channel outputs and conferencing messages received during the Rx-conferencing phase. 
\end{enumerate}

\bigskip

\begin{remark}
In the above problem formulation, the \emph{Tx-conferencing phase} takes place before the \emph{communications phase}, and the \emph{Rx-conferencing phase} takes place before the \emph{decoding phase}. The conferencing phases, however, should not be considered as strictly separated from the communications and decoding phases. In fact, a transmitter might compute and store its transmit signal already during the Tx-conferencing phase. Similarly, a receiver might compute and store its decoded message already during the Rx-conferencing phase. This allows transmitters (resp. receivers) to exchange parts of their transmit signals (resp. decoded messages) over the conferencing links with their neighbours. 
\end{remark}

\bigskip

We now describe the four communication phases more formally.

\subsection{Tx-Conferencing Phase with $\kappaT$ Rounds}

To model systems with tight latency, computational complexity or storage space constraints, we shall focus on the case where the number of Tx- and Rx-conferencing rounds are limited to (finite) $\kappaT \in \nintegers$ and $\kappaR \in \nintegers$ respectively. (Later, for comparison, we will also consider the case of unlimited conferencing rounds.)

The Tx-conferencing phases consists of $\kappaT$ rounds: In round $j = 1,2,\ldots,\kappaT$, transmitter~$k$ sends a conferencing message $U_{k\to k'}\bs{j}$ to  its neighbouring transmitters $k-1$ and $k+1$. Here,
\begin{equation*}
U_{k\to k'}\bs{j} := \phi_{k,k'}^{(j)}\Big( M_k\ U\bs{1}_{k-1 \to k}\ U\bs{1}_{k+1 \to k}\ U\bs{2}_{k-1 \to k}\ U\bs{2}_{k+1 \to k}\ \ldots\ U\bs{j-1}_{k-1 \to k}\ U\bs{j-1}_{k+1 \to k}\Big),
\end{equation*} 
where
\begin{equation}\label{Eqn:EncoderConf}
\phi_{k,k'}\bs{j} : \mathcal{M}_k \times 
\prod_{j'=1}^{j-1} 
\prod_{\tilde{k} \in \{k-1,k+1\}}
\sRTx{j'}{\tilde{k}}{k}
\longrightarrow 
\sRTx{j}{k}{k'}
\end{equation}
and we understand $U\bs{j}_{0 \to 1}$ and $U\bs{j}_{K+1 \to K}$ to be degenerate random variables (constants) with 
\begin{equation*}
\RTx{j}{0}{1} = \RTx{j}{K+1}{K} = 0, \quad \forall j\in\{1,\ldots, \kappaT\}.
\end{equation*}
The total rate from transmitter $k$ to each one of its neighbours may not exceed the rate budget:
\begin{equation}\label{eq:txconferencing_constraint}
\sum_{j = 1}^\kappaT \RTx{j}{k}{k'} \leq \RT
\end{equation}
for all $k' \in \{k-1, k+1\}$ and $k \in \sK$. 
 

\subsection{Cooperative-Communication Phase}
The channel codeword sent by transmitter $k$, 
\begin{equation*}
X_k^n = (X_{k,1}\ X_{k,2}\ \ldots\ X_{k,n}),
\end{equation*}
is a function of its source message $M_k$ and the conferencing messages it received during the Tx-conferencing phase. Specifically
\begin{equation*}
X_k^n  := \phi_{k,k'}^{(j)}\Big( M_k\ U\bs{1}_{k-1 \to k}\ U\bs{1}_{k+1 \to k}\ U\bs{2}_{k-1 \to k}\ U\bs{2}_{k+1 \to k}\ \ldots\ U\bs{\kappaT}_{k-1 \to k}\ U\bs{\kappaT}_{k+1 \to k}\Big),
\end{equation*} 
and
\begin{equation}\label{Eqn:Encoder}
f_k : \mathcal{M}_k \times  
\prod_{j = 1}^{\kappaT}\
\prod_{k' \in \{k-1,k+1\}}
\sRTx{j}{k'}{k}
\rightarrow \mathbb{R}^n.
\end{equation}


\subsection{Rx-Conferencing Phase with $\kappaR$ Rounds}
The Rx-conferencing phase takes place after all the channel outputs have been observed by the receivers. Let  
\begin{equation*}
Y_k^n = (Y_{k,1}\ Y_{k,2}\ \ldots\ Y_{k,n})
\end{equation*}
denote the channel outputs observed at receiver $k$. 

The Rx-conferencing phase consists of $\kappaR$ rounds: In round $j$, receiver~$k$ sends a conferencing message $V_{k\to k'}\bs{j}$ to its neighbour receiver~$k'$, for  $k'\in\{k-1,k+1\}$. Here 
\begin{equation*}
V_{k\to k'}\bs{j} := \phi_{k,k'}\bs{j}\Big( Y_k^n\ V\bs{1}_{k-1 \to k}\ V\bs{1}_{k+1 \to k}\ V\bs{2}_{k-1 \to k}\ V\bs{2}_{k+1 \to k}\ \ldots\ V\bs{j-1}_{k-1 \to k}\ V\bs{j-1}_{k+1 \to k}\Big),
\end{equation*} 
where
\begin{equation}\label{Eqn:DecoderConf}
\psi_{k,k'}^{(j)} : 
\mathbb{R}^n 
\times 
\prod_{j'=1}^{j-1}
\prod_{\tilde{k}\in\{k-1,k+1\}} 
\sRRx{j'}{\tilde{k}}{k}
\longrightarrow 
\sRRx{j}{k}{k'}
\end{equation}
and we understand $V\bs{j}_{0 \to 1}$ and $V\bs{j}_{K+1 \to K}$ to be degenerate random variables (constants) with 
\begin{equation*}
\RRx{j}{0}{1} = \RRx{j}{K+1}{K} = 0, \quad \forall j\in\{1,\ldots, \kappaR\}.
\end{equation*}
We require that the total rate from receiver~$k$ to its immediate neighbours does not exceed the rate budget, 
\begin{equation} \label{eq:rxconferencing_constraint}
\sum_{j = 1}^\kappaR \RRj{j} \leq \RR, \qquad \forall\ k' \in\{k-1,k+1\}.
\end{equation}


\subsection{Decoding Phase}

Receiver $k$ estimates the source message $M_k$ by
\begin{equation*}
\h{M}_{k} := g_k\bs{j}\Big( Y_k^n\ V\bs{1}_{k-1 \to k}\ V\bs{1}_{k+1 \to k}\ V\bs{2}_{k-1 \to k}\ V\bs{2}_{k+1 \to k}\ \ldots\ V\bs{\kappaR}_{k-1 \to k}\ V\bs{\kappaR}_{k+1 \to k}\Big)
\end{equation*}
where
\begin{equation}\label{Eqn:Decoder}
g_k : \mathbb{R}^n 
\times 
\prod_{j' = 1}^{\kappaR}
\prod_{k' \in \{k-1,k+1\}} 
\sRRx{j'}{k'}{k}
\rightarrow 
\mathcal{M}_k. 
\end{equation}


\subsection{Capacity, Sum Capacity and Per-User Multiplexing Gain}

For brevity, let 
\begin{equation*}
\bfkappa = (\kappaT\ \kappaR),\quad \bfmu = (\muT\ \muR)\;\; \text{ and } \;\; \bfR = (R_1\ \ldots\ R_K).
\end{equation*}
We call the collection of encoders and decoders in~\eqref{Eqn:EncoderConf}, \eqref{Eqn:Encoder}, \eqref{Eqn:DecoderConf} and~\eqref{Eqn:Decoder} an $(n,\bfR,\bfkappa,\bfmu,P)$-\emph{code}. For given $\bfkappa$, $\bfmu$ and $P$: We say that a rate tuple $\bfR$ is $(\bfkappa, \bfmu, P)$-\emph{achievable} if for every $\epsilon > 0$ there exists a sufficiently large blocklength~$n$ and an $(n,\bfR,\bfkappa,\bfmu,P)$-code with 
\begin{equation*}
\Pr \big[(\h{M}_1\ \ldots\ \h{M}_K) \neq (M_1\ \ldots\ M_K)\big] \leq \epsilon.
\end{equation*}
The \emph{capacity region} $\Ca(\bfkappa,\bfmu,P)$ is the closure of the set of all $(\bfkappa,\bfmu,P)$-achievable rate tuples, and the \emph{sum capacity} is 
\begin{equation*}
\Cs(\bfkappa,\bfmu,P) := \max_{\bfR \in \Ca(\bfkappa,\bfmu,P)}\ \sum_{k=1}^{K} R_k.
\end{equation*}

\bigskip

\begin{definition}
The \emph{per-user multiplexing gain (MG)} is 
\begin{equation*}
\S(\bfkappa,\bfmu)
:= \varlimsup_{K\rightarrow \infty}\; \varlimsup_{P\rightarrow\infty} \;\frac{\Cs(\bfkappa,\bfmu,P)}{K\cdot\frac{1}{2}\log P},
\end{equation*}
for $\bfkappa \in \nintegers \times \nintegers$ and $\bfmu \in [0,\infty) \times [0,\infty)$. 
\end{definition}

\bigskip

The main problem of interest in this paper is to determine $\S(\bfkappa,\bfmu)$. In Section~\ref{Sec:MainResults}, we give upper and lower bounds on $\S(\bfkappa,\bfmu)$ and show that these bounds coincide in many cases. The next proposition summarises some basic properties of the per-user MG, and we omit its proof. 
 
\bigskip

\begin{proposition}\label{Prop:MG}
\begin{enumerate}

\item $\S(\bfkappa,\bfmu)$ is nondecreasing in $\bfkappa$ and $\bfmu$ and upper bounded by $1$. 

\item If $\muT = 0$, then $\S(\bfkappa,\bfmu)$ does not depend on $\kappaT$:
\begin{equation*}
\S(0,\kappaR,0,\muR) = \S(1,\kappaR,0,\muR) = \cdots
\end{equation*}
for all $\kappaR \in \nintegers$ and $\muR \in [0,\infty)$. \\Similarly, if $\muR = 0$, then  $\S(\bfkappa,\bfmu)$ does not depend on $\kappaR$:
\begin{equation*}
\S(\kappaT,0,\muT,0) = \S(\kappaT,1,\muT,0) = \cdots
\end{equation*}
for all $\kappaT \in \nintegers$ and $\muT \in [0,\infty)$. 

\item If $\bfmu = (0,0)$, then 
\begin{equation*}
\S(\kappaT,\kappaR,0,0) = \frac{1}{2}
\end{equation*}
for all $(\kappaT,\kappaR) \in \nintegers \times \nintegers$.
\end{enumerate}
\end{proposition}

 
\subsection{Conferencing with Unlimited Rounds}

To help put our results $\S(\bfkappa,\bfmu)$ in context, we will also consider the case of unlimited Tx- and Rx-conferencing rounds. The \emph{per-user MG with unlimited conferencing} is defined in the same way as above, except now $\kappaT$ and $\kappaR$ are infinite\footnote{For a given blocklength $n$ scheme, it however suffices to choose $\kappaT\leq 2 K n R_\Tx$ and $\kappaR \leq 2 K n R_\Rx$, because each of the $K$ transmitters can send at most $n R_\Tx$ bits over each of the two links to its left and right, and similarly each of the $K$ receivers can send at most $n R_\Rx$ bits over each of the two links to its left and right.}.  Let $\S_{\infty}(\bfmu)$ denote the per-user MG with unlimited conferencing. The next proposition summarises some basic properties of $\S_\infty(\bfmu)$, and we omit its proof.
  
\bigskip

\begin{proposition}\label{Prop:UnlimitedMG}
\begin{enumerate}

\item $\S_{\infty}(\bfmu)$ is nondecreasing in $\bfmu$.

\item For all $\bfkappa \in \nintegers \times \nintegers$ and $\bfmu \in [0,\infty) \times [0,\infty)$:
\begin{equation*}
\frac{1}{2} \leq \S(\bfkappa,\bfmu) \leq \S_{\infty}(\bfmu) \leq 1
\end{equation*}

\end{enumerate}
\end{proposition}


\section{Main Results}\label{Sec:MainResults}


\subsection{Converse: Conferencing with Finite Rounds}

Our first theorem gives a converse (upper) bound on the per-user MG $\S(\bfkappa,\bfmu)$. 

\bigskip

\begin{theorem}\label{Thm:Converse}
For all $\bfkappa = (\kappaT\ \kappaR) \in \nintegers \times \nintegers$ and $\bfmu = (\muT\ \muR) \in [0,\infty) \times [0,\infty)$, we have
\begin{equation}\label{eq:convres1}
\S(\bfkappa,\bfmu) 
\leq 
\min\left\{\frac{2\muT + 2\muR + 1}{2}\, ,\,\frac{ 2 \kappaT+2 \kappaR+1}{2 \kappaT+2\kappaR+2} \right\}.
\end{equation}
If $\muT=0$, then
\begin{equation}\label{eq:convres2}
 \S(\bfkappa, \bfmu) \leq \min\left\{ \frac{2\muR + 1}{2}\, ,\,\frac{2\kappaR+1}{2\kappaR+2} \right\}.
\end{equation}
If $\muR=0$, then
\begin{equation}\label{eq:convres3}
 \S(\bfkappa,\bfmu) \leq \min\left\{\frac{2\muT + 1}{2}\, ,\,\frac{ 2 \kappaT+1}{2 \kappaT+2} \right\}.
\end{equation}
\end{theorem}

\bigskip

\begin{IEEEproof}
Theorem~\ref{Thm:Converse} is proved in Section~\ref{Sec:Proof:Thm:Converse}.
\end{IEEEproof}


\subsection{Achievability: Conferencing with Finite Rounds}

The next theorem gives an achievable (lower) bound on the per-user MG $\S(\bfkappa,\bfmu)$. Its expression depends on the following two quantities
\begin{equation}\label{eq:quotients}
\pi_\Tx  := \frac{\muT}{\kappaT} 
\quad 
\text{and}
\quad
\pi_\Rx  := \frac{\muR}{\kappaR}.
\end{equation}

Define $\Sach: \pintegers \times \pintegers \times [0,\infty) \times [0,\infty) \to [0,1]$ as follows:
\begin{itemize} 
\begin{subequations}\label{eq:Sach}

\item If $\pi_\Rx = \pi_\Tx$, let
\begin{equation}
\Sach(\bfkappa,\bfmu) := 
\begin{cases} 
\dfrac{2 \muT + 2\muR + 1}{2}
&\text{if}\quad 
2 \muT + 2\muR + 2\pi_{\Tx} \leq 1 \\[12pt] 
\dfrac{2 \kappaT+2\kappaR+1}{2\kappaT+2\kappaR+2} 
&\text{otherwise.}
\end{cases} 
\end{equation}

\item If $\pi_\Rx < \pi_\Tx$, let  
\begin{equation}\label{eqn:thm:aa:case2}
\Sach(\bfkappa,\bfmu) := 
\begin{cases} 
\dfrac{2\muT + 2\muR + 1}{2} 
&\text{if}\quad 
2\muT + 2\muR + 2\pi_{\Tx} \leq 1\\[12pt] 
\dfrac{2 \kappaT+2\kappaR+1}{2\kappaT+2\kappaR+2} 
&\text{if}\quad 
2\muT \left(\dfrac{\pi_{\Rx}}{\pi_{\Tx}}\right) + 2\muR + 2\pi_{\Rx}  > 1\\[12pt] 
\dfrac{2\kappaT + 2\muR + 1}{2\kappaT+2} 
& \textnormal{otherwise}.
\end{cases} 
\end{equation}

\item If $\pi_\Rx > \pi_\Tx$, let  
\begin{equation}
\Sach(\bfkappa,\bfmu) := 
\begin{cases} 
\dfrac{2\muT + 2\muR + 1}{2} 
&\text{if}\quad 
2\muT+2\muR+ 2\pi_{\Rx} \leq 1\\[12pt] 
\dfrac{2 \kappaT+2\kappaR+1}{2\kappaT + 2\kappaR + 2}
&\text{if}\quad 
2\muT + 2\muR \left(\dfrac{\pi_\Tx}{\pi_\Rx}\right) + 2\pi_{\Tx}  > 1\\[12pt] 
\dfrac{2\kappaR + 2\muT + 1}{2\kappaR + 2}
&\textnormal{otherwise}.
\end{cases} 
\end{equation}
\end{subequations}
\end{itemize}

\bigskip

\begin{theorem}\label{Thm:Ach}
For all $\bfkappa \in \pintegers \times \pintegers$ and $\bfmu \in [0,\infty) \times [0,\infty)$, we have
\begin{equation*}
\S(\bfkappa,\bfmu)\geq \Sach(\bfkappa,\bfmu).
\end{equation*}
\end{theorem}

\bigskip

\begin{IEEEproof}
Theorem~\ref{Thm:Ach} is proved in Section~\ref{Sec:Proof:Thm:Ach}.
\end{IEEEproof}
\bigskip
\begin{remark} 
The coding scheme that we present in Section~\ref{Sec:Proof:Thm:Ach} requires that the codebooks are known during the Tx- and Rx-conferencing phases. When $\kappaT=\kappaR=1$, then it is  possible to find a coding scheme achieving $\Sach(\bfkappa,\bfmu)$ where the conferencing phases do not use knowledge about codebooks.  Our results for  $\kappaT=\kappaR=1$ thus continue to hold also in  oblivious setups.  The scheme in Section~\ref{Sec:Proof:Thm:Ach} needs to be changed as follows: Instead of sending quantised versions of transmit signals over the Tx-conferencing links, the transmitters conference  source messages, and instead of sending decoded sources messages over the Rx-conferencing links, the receivers send quantised versions of their receive signals. (Details omitted.)
\end{remark} 


\subsection{Conferencing with Unlimited Number of Rounds}

Now consider the case where the number of conferencing rounds is unconstrained. The next theorem determines the exact per-user MG.  

\bigskip

\begin{theorem}\label{Thm:UnlimitedConf} 
\begin{equation}
\S_{\infty}(\bfmu)  
=\min\left\{1\, , \, \frac{1+ 2 \muT+2\muR}{2}\right\}
\end{equation}
for all $\bfmu \in [0,\infty) \times [0,\infty)$.
\end{theorem}
 
\bigskip 
 
\begin{IEEEproof}
The proof of Theorem~\ref{Thm:UnlimitedConf} consists of a direct part proving
\begin{equation}
\S_{\infty}(\bfmu)  
\geq \min\left\{1\, , \, \frac{1+ 2 \muT+2\muR}{2}\right\}
\end{equation}
and a converse part proving
\begin{equation}\label{eq:converse_infinity}
\S_{\infty}(\bfmu)  
\leq\min\left\{1\, , \, \frac{1+ 2 \muT+2\muR}{2}\right\}.
\end{equation}
The converse part is proved in Section~\ref{sec:converse_unlimited}.

The direct part follows from Theorem~\ref{Thm:Ach} and  from continuity considerations. Specifically, for
\begin{equation}\muT+\muR<\frac{1}{2},
\end{equation} the desired per-user MG of  $\frac{1+ 2 \muT+2\muR}{2}$ is achievable by Theorem~\ref{Thm:Ach} when one chooses the number of conferencing rounds $\kappaT$ and $\kappaR$ sufficiently large so that $2 \muT +2\muR +2 \min\Big\{\frac{\muT}{\kappa_\Tx}, \, \frac{\muR}{\kappa_\Rx} \Big\}<1$. (Since here the number of conferencing rounds is unlimited, we can choose $\kappa_\Tx$ and $\kappa_\Rx$ as large as we wish.)
Moreover, since the per-user MG is non-decreasing in   $\muT$ and $\muR$ (see Proposition~\ref{Prop:MG})  a per-user MG of $1$ must be achievable whenever  
\begin{equation*}
\muT+\muR\geq \frac{1}{2}.
\end{equation*}

\end{IEEEproof} 
 

\subsection{Discussion and Corollaries to Theorems~\ref{Thm:Converse},~\ref{Thm:Ach} and~\ref{Thm:UnlimitedConf}}

\begin{figure}[t!]
\centering
 \includegraphics[height=0.3 \textheight]{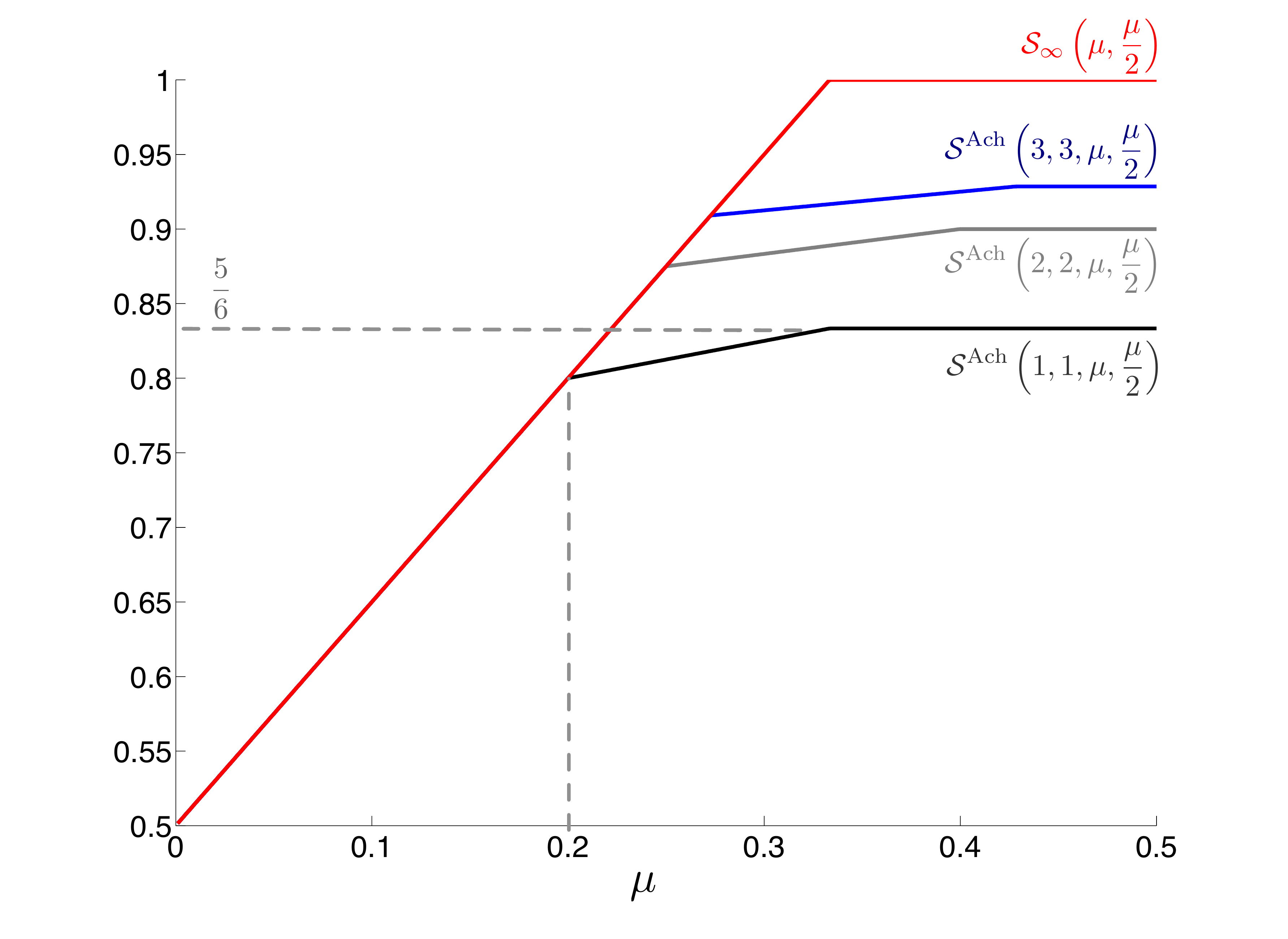}
 \caption{The achievable (lower) bound $\Sach(\bfkappa,\bfmu)$ in \eqref{eq:Sach} on the per-user MG $\S(\bfkappa,\bfmu)$ for the case with equal Tx- and Rx-conferencing rounds $\bfkappa=(\kappa,\kappa)$ and unbalanced Tx and Rx prelog conferencing contraints $\muT=2\muR$. The asymptotic per-user MG for unlimited conferencing rounds $\S_\infty(\bfmu)$, as described in Theorem~\ref{Thm:UnlimitedConf}, is also shown.}
  \label{fig:prelog6}
\end{figure}
We start with an example.
\begin{example}\label{exa}
Let
\begin{equation*}
\bfkappa = (\kappa,\kappa) 
\text{ and }
\bfmu = (\mu,\mu/2).
\end{equation*}
In this case, the achievable bound from Theorem~\ref{Thm:Ach}, $\Sach(\kappa,\kappa,\mu,\mu / 2)$, is given by~\eqref{eqn:thm:aa:case2}. Figure~\ref{fig:prelog6} plots~\eqref{eqn:thm:aa:case2} as a function of $\mu$ for $\kappa = 1,2$ and $3$ conferencing rounds. The figure also shows the per-user MG with unlimited conferencing rounds from Theorem~\ref{Thm:UnlimitedConf},  
\begin{equation*}
\S_\infty\left(\mu,\frac{\mu}{2}\right)
= 
\left\{
\begin{array}{cl}
\dfrac{3 \mu + 1}{2} & \text{ if } 0 \leq \mu \leq \dfrac{1}{3}\\[10pt]
1 &\text{otherwise}.
\end{array}
\right.
\end{equation*}

If the prelog constant $\mu$ is small, then it can be seen from Figure~\ref{fig:prelog6} that a single conferencing round achieves the same per-user MG as multiple (even unlimited) rounds; that is, for all $\mu \in [0,1/5]$
\begin{equation*}
\S_\infty \left(\mu, \frac{\mu}{2}\right) = \Sach\left(1,1,\mu,\frac{\mu}{2}\right) = \Sach\left(2,2,\mu,\frac{\mu}{2}\right) = \ldots.
\end{equation*}
\end{example}
\bigskip
This idea extends to more general setups: Finitely many conferencing rounds are optimal whenever the prelog conferencing constants are sufficiently small. 

\begin{figure}[t]
\begin{center}
\includegraphics[width=0.6\textwidth]{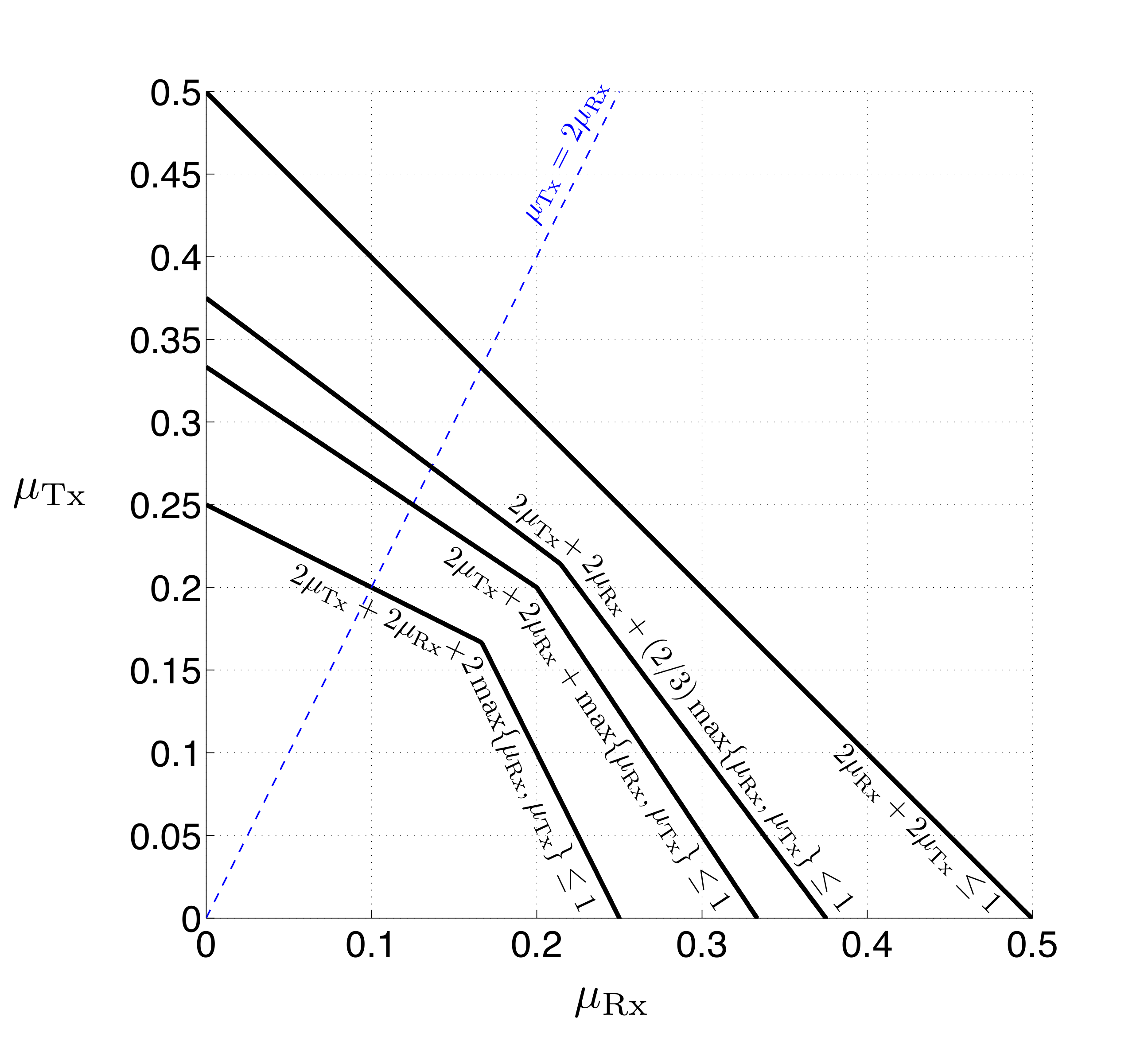}
\caption{An illustration of the \emph{small $\bfmu$ regions} of Corollary~\ref{Cor:SmallMu}, where a single, two or three conferencing rounds suffice. The topmost line indicates the values of $\muT, \muR$ where the per-user MG saturates at 1 in the case of unlimited conferencing rounds.}
\label{Fig:SmallMu}
\end{center}
\end{figure}

\bigskip

\begin{corollary}[Small $\bfmu$ regions]\label{Cor:SmallMu}
Fix the number of Tx- and Rx-conferencing rounds $\bfkappa=(\kappaT,\kappaR)$. If the prelog conferencing constants $\bfmu=(\muT,\muR)$ are  small (depending on $\bfkappa$), namely,
\begin{equation*}
2 \muT+2\muR+ 2\max\left\{ \frac{\muT}{\kappaT},\frac{\muR}{\kappaR}\right\}\leq 1,
\end{equation*}
then $\bfkappa$ conferencing rounds achieve the same performance as unlimited conferencing rounds; that is,
\begin{equation}
\S(\bfkappa,\bfmu) = \S(\bfkappa',\bfmu)  = \S_\infty(\bfmu) = \frac{1 + 2 \muT+2\muR}{2}
\end{equation}
for all $\bfkappa' \in \pintegers \times \pintegers$ with $\kappaT \leq \kappaT'$ and $\kappaR \leq \kappaR'$.
\end{corollary}

\begin{IEEEproof}
Corollary~\ref{Cor:SmallMu} follows directly from Theorems~\ref{Thm:Converse} and~\ref{Thm:Ach}.
\end{IEEEproof}

\bigskip

Figure~\ref{Fig:SmallMu} illustrates  the regions of small $\bfmu$s indicated in Corollary~\ref{Cor:SmallMu} where a small number of conferencing rounds yields the same per-user MG as an unlimited number of rounds. 
Specifically, the first line ($2\muT + 2\muR + \max\{\muT,\muR\} \leq 1$) depicts the boundary of the small $\bfmu$ region where one-shot conferencing ($\kappa = 1$) is optimal, i.e., achieves the same per-user MG as unlimited-rounds conferencing. 
The second line ($2\muT + 2\muR + \max\{\muT,\muR\} \leq 1$) depicts the boundary of the small $\bfmu$ region where two-round conferencing ($\kappa = 2$) is optimal, and the third line ($2\muT + 2\muR + (2/3)\max\{\muT,\muR\} \leq 1$) depicts the boundary of the small $\bfmu$ region where three-round conferencing ($\kappa = 3$) is optimal. The top-most line depicts the $\bfmu$ values for which in case of unlimited conferencing rounds the per-user MG saturates at 1. The blue dashed line  ($\muT = 2\muR = \mu$) corresponds to the example in figure~\ref{fig:prelog6}: its crossing point with the  $2\muT + 2\muR + \max\{\muT,\muR\} \leq 1$-line, for example, shows that $\kappa = 1$ is optimal for all $\mu \in [0,1/5]$.

\bigskip
We now switch to the  \emph{large $\bfmu$ regime}. Consider again example~\ref{exa} and    Figure~\ref{fig:prelog6}. For $\mu > \kappa / (2\kappa + 1)$, where the achievable lower bound 
in~\eqref{eqn:thm:aa:case2} meets the upper bound in Theorem~\ref{Thm:Converse} and therefore
\begin{equation*}
\S\left(\kappa,\kappa,\mu,\frac{\mu}{2}\right) = \frac{4\kappa + 1}{4\kappa + 2}.
\end{equation*}
We can see that in this regime $\S(\kappa,\kappa,\mu,\mu/2)$ is saturated in terms of conferencing prelog $\mu$, but is  strictly increasing in $\kappa$, so adding additional conferencing rounds increases the per-user MG. 

More generally, 
the achievable bound  in Theorem~\ref{Thm:Ach} is optimal whenever $\bfmu$ is sufficiently large. The resulting expression for $\Sach(\bfkappa,\bfmu)$ moreover exhibits that  in this large-$\bfmu$ regime the per-user MG saturates in $\bfmu$ but is strictly increasing in the maximum number of allowed conferencing rounds $\kappaT$ and $\kappaR$.
\bigskip
\begin{corollary}\label{Cor:LargeMu}
Fix the number of Tx- and Rx- conferencing rounds $\bfkappa = (\kappaT\ \kappaR)$. If the prelog conferencing constants $\bfmu = (\muT\ \muR)$ are sufficiently large (depending on $\bfkappa$), namely, 
\begin{equation*}
\muT\cdot \min\left\{1,\frac{\piR}{\piT} \right\}
+ \muR \cdot\min\left\{1,\frac{\piT}{\piR} \right\}
+ \min\big\{ \piT,\piR \big\}
> 
\frac{1}{2},
\end{equation*}
then,
\begin{equation}\label{eqn:limMuSach}
\S(\bfkappa,\bfmu) =
 \frac{2 \kappaT+2\kappaR+1}{2\kappaT+2\kappaR+2}.
\end{equation}
\end{corollary} 

\bigskip

Intuitively, the right hand side of~\eqref{eqn:limMuSach} is smaller than one---irrespective of the available prelogs on the conferencing links---because the interference cancelation techniques that we employ in our coding scheme (see Section~\ref{Sec:Proof:Thm:Ach}) cause interference to propagate through the network. This ``propagating interference" can be (partially) eliminated only at close Tx/Rx pairs within the range of conferencing. For example, a conference message detailing the interference caused by transmitter $1$ can be relayed from transmitter $1$ over the Tx-conferencing links as far as transmitter $\kappaT + 1$, but no further. Our scheme stops interference propagating beyond the conferencing range by selectively ``shutting down'' transmitters, and these shutdowns cause $\Sach(\bfkappa,\bfmu)$ to be strictly less than $1$. 

\bigskip
Our achievable bound $\Sach(\bfkappa,\bfmu)$ in Theorem~\ref{Thm:Ach} is also optimal when there is conferencing at the transmitters or receivers, but not both. 

\bigskip

\begin{corollary}\label{cor:e}
\begin{itemize}
\item[] If $\muT=0$, then 
\begin{equation}
\S(\bfkappa,\bfmu)=
\begin{cases} 
\dfrac{1+2\muR}{2}, &  2\muR+ 2 \pi_{\Rx} \leq 1 \\[10pt]  
\dfrac{2 \kappaR+1}{2\kappaR+2}, & \text{otherwise}
\end{cases} 
\end{equation}
for all $\bfkappa \in \pintegers \times \pintegers$ and $\muR \in [0,\infty)$.

\item If $\muR=0$, then
\begin{equation}
\S(\bfkappa,\bfmu) = 
\begin{cases} 
\dfrac{1+2\muT}{2}, &  2\muT+ 2 \pi_{\Tx} \leq 1 \\[10pt] 
\dfrac{2 \kappaT+1}{2\kappaT+2}, & \textnormal{ otherwise}.
\end{cases} 
\end{equation}
for all $\bfkappa \in \pintegers \times \pintegers$ and $\muT \in [0,\infty)$.
\end{itemize}
\end{corollary}

\bigskip
For the case of transmitter-conferencing only ($\muR=0$), we observe that: 
\begin{itemize} 
\item A single conferencing round is optimal whenever $\muT\leq \frac{1}{4}$.
\item Two conferencing rounds are optimal whenever $\muT \leq \frac{1}{3}$.
\item Three conferencing rounds are optimal whenever $\muT\leq \frac{3}{8}$.
\item Generally, $\kappa\in\mathbb{Z}^+$  conferencing rounds are optimal whenever $\muT\leq \frac{\kappa}{2(\kappa+1)}$.\\
\end{itemize}

Finally, the achievable bound $\Sach(\bfkappa,\bfmu)$ is also tight under certain symmetry conditions. 

\bigskip

\begin{corollary}\label{Cor:Sym}
If $\pi_\Tx=\pi_\Rx=\pi$, then
\begin{equation}
 \S(\bfkappa,\bfmu) =\begin{cases} 
\dfrac{1+ 2 \muT+2\muR}{2},&
2 \muT+2\muR+ 2 \pi \leq 1 \\[10pt] 
\dfrac{2 \kappaT+2\kappaR+1}{2\kappaT+2\kappaR+2}, 
&\text{otherwise}.
\end{cases} 
\end{equation}
\end{corollary}


%


\section{Proof of Theorem~\ref{Thm:Ach}}\label{Sec:Proof:Thm:Ach}


\subsection{Overview}

\begin{figure}
\begin{center}
\includegraphics[width=\textwidth]{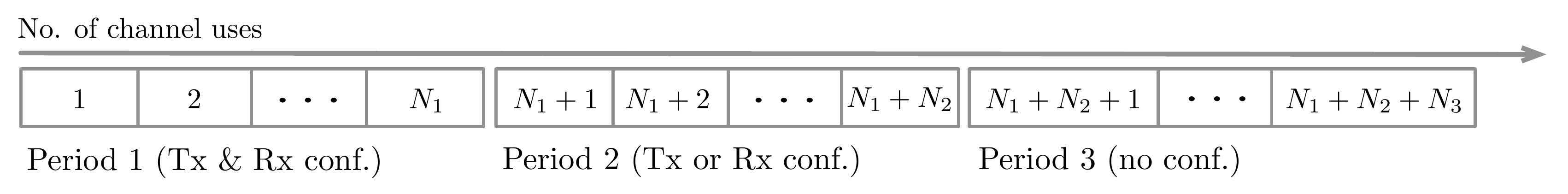}
\caption{Timesharing between Tx and Rx conferencing schemes}
\label{Fig:Timesharing}
\end{center}
\end{figure}

We now present a coding strategy that achieves $\Sach(\bfkappa,\bfmu)$. The strategy will time-share between one, two or three different coding schemes --- depending on the particular values of $\bfkappa$ and $\bfmu$. To this end, let us divide the blocklength $n$ in three consecutive periods of lengths $N_1$, $N_2$ and $N_3$ channel symbols, as shown in Fig.~\ref{Fig:Timesharing}, so that $n = N_1 + N_2 + N_3$. Here $n$ and, therefore, $N_1$, $N_2$ and $N_3$ can be chosen arbitrarily large. 
\begin{itemize}

\item \emph{Period 1:} During period 1 (the first $N_1$ channel uses) we will use a scheme that employs both Tx- and Rx-conferencing, assuming that the prelogs $\muT$ and $\muR$ are both positive. If $\muT = 0$ or $\muR = 0$, then we will remove period 1 by setting $N_1 = 0$. 

\item \emph{Period 2:} During period 2 (channel uses $N_1 + 1$ to $N_1 + N_2$) we will use a scheme that employs either Tx-conferencing (when $\pi_\Rx<\pi_\Tx$) or Rx-conferencing (when $\pi_\Tx > \pi_\Rx$), but not both. If $\pi_\Rx =\pi_\Tx$, then we will remove period 2 by setting $N_2 = 0$. 

\item \emph{Period 3:} During period 3 (the last $N_3$ channel uses) we will use a scheme that does not employ Tx- or Rx-conferencing. 

\end{itemize}


We now choose $N_1$, $N_2$ and $N_3$, and detail the coding schemes used during each period. To simplify exposition, suppose that 
\begin{equation} \label{eq:assumption}
\pi_\Tx \geq \pi_\Rx,
\end{equation}
so that we only use Tx-conferencing during period 2. (The case $\pi_\Tx < \pi_\Rx$ can be treated by exchanging the subscripts $\Tx$ and $\Rx$ everywhere in the following arguments.) Choose
\begin{subequations}\label{eq:choiceN}
\begin{align}
N_1 & := \min\big\{ n,\ n\piR(2\kappaT + 2\kappaR + 2) \big\}\\
N_2 & := \min\big\{ n - N_1,\ n(\piT- \piR)(2\kappaT+2)\big\}\\
N_3 & := n-N_1-N_2.
\end{align}
\end{subequations}

\emph{Period 1:} We will timeshare between $2 \kappaT+2 \kappaR+2$ instances of the coding scheme described in Section~\ref{sec:tx_rx_conferencing} but where we make sure that each (sub)message is sent at a rate not exceeding $\frac{1}{2} \log (1+P)$.\footnote{This ensures that also each Rx-conferencing link is used at a rate of at most $\frac{1}{2} \log(1+P)$.}  Each of these instances is used over an equally long interval, i.e., over $\lfloor N_1/(2 \kappaT+2 \kappaR+2)\rfloor$ channel symbols, and employs a different value of the parameter $\i\in\{1, \ldots, 2 \kappaT+2\kappaR+2\}$. Varying the parameter $\i$ over $1, \ldots, 2 \kappaT+2\kappaR+2$ varies the utilised resources (transmit powers and conferencing links) as well as the served Tx/Rx pairs in a round robin manner.

\emph{Period 2:} We will timeshare between $2\kappaT+2$ instances of the coding scheme in Section~\ref{sec:tx_conferencing}, where we again make sure that each message is sent at a rate not exceeding $\frac{1}{2}\log(1+P)$.\footnote{This again ensures that each Rx-conferencing link is used at a rate of at most $\frac{1}{2} \log(1+P)$.} Each of these instances is again used over an equally-long interval, i.e., over $\lfloor N_2/(2 \kappaT+2)\rfloor$ channel uses, and employs a different value for the parameter $\i=1,\ldots, 2\kappaT+2$. 

\emph{Period 3} We will employ the scheme in Section~\ref{sec:no_conferencing}. No conferencing is used. 

We now present the coding schemes that we time-share in the various transmission periods.


\subsection{Tx- and Rx-Conferencing in Period 1}\label{sec:tx_rx_conferencing} 


Let
\begin{equation*}
\beta  := 2 \kappaT+2 \kappa_\Rx+2.
\end{equation*}
Choose $\i\in\{1,\ldots, \beta\}$ arbitrarily, and define 
\begin{equation*}
\gamma := \left\lfloor\frac{ K-\i+1}{\beta} \right\rfloor. 
\end{equation*}


\subsubsection{Split the Network into Subnetworks}

\begin{figure}
\begin{center}
\includegraphics[width=\textwidth]{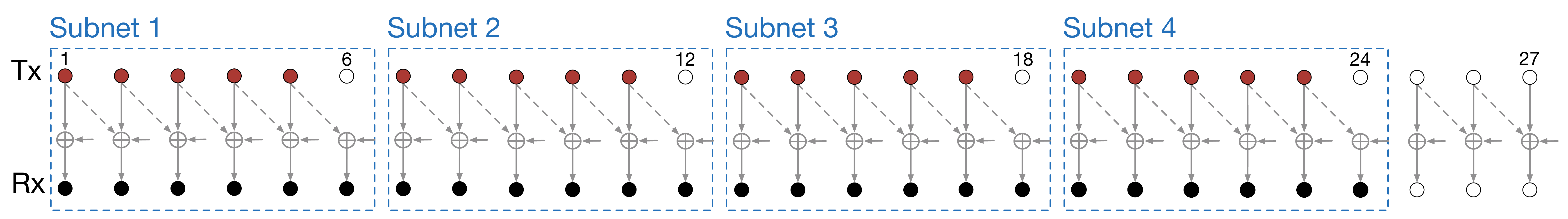}
\caption{Transmitter and receiver conferencing scheme: The network is decomposed into $\gamma$ non-interfering subnets, where each subnet consists of $(\beta-1)$ active transmitters and $\beta$ active receivers. Red nodes represent active transmitters, black nodes represent active receivers, and white nodes represent deactivated transmitters. Example parameters: $K = 27$, $\i = 1$, $\beta = 6$ and $\gamma = 4$.}
\label{Fig:Split_Network_Mode2_V2}
\end{center}
\end{figure}

We first split the network into $\gamma$ identical subnets that do not interfere each other. To this end, we will deactivate (silence) every transmitter\footnote{Transmitters~$1, \ldots, \i-1$ and~$\i+ \gamma \beta,\ldots, K$ have been deactivated to simplify the following presentation. In fact, we could improve the scheme's performance by, for example, reactivating transmitters $1,3,5,\ldots,i-1$ (each reactivated transmitter can communicate with its receiver over an interference-free Gaussian point-to-point channel). Such reactivations, however, will not improve the scheme's asymptotic ($K \to \infty$) per-user MG.} with an index 
\begin{equation*}
k \in \mathcal{S}:=\{1,\ldots, \i-1\} \cup \{\i+\beta-1, \i+2 \beta-1, \ldots, \i+\gamma \beta-1\} \cup \{\i+ \gamma \beta,\ldots, K\}.
\end{equation*}
That means, every silenced transmitter (Tx)~$k\in\mathcal{S}$ sets its channel inputs $X_k^n$ deterministically to 0. Moreover, every such silenced Tx~$k\in\mathcal{S}$ can send and receive conferencing messages only to and from its left-neighbour Tx~$k-1$, but not  its right-neighbour Tx~$k+1$. Similarly, each corresponding receiver (Rx) $k\in\s{S}$ can send and receive conferencing messages only to and from  its left-neighbour Rx~$k-1$, but not its right-neighbour Rx~$k+1$. According to these assumptions, the various subnets do not interfere and they each consist of $(\beta-1)$ active transmitters and $\beta$ active receivers --- an example is illustrated in Figure~\ref{Fig:Split_Network_Mode2_V2}.

\subsubsection{Communication in a Subnet}\label{sec:subnetcom}
\begin{figure}[h!]
\begin{center}
\includegraphics[width=0.925\textwidth]{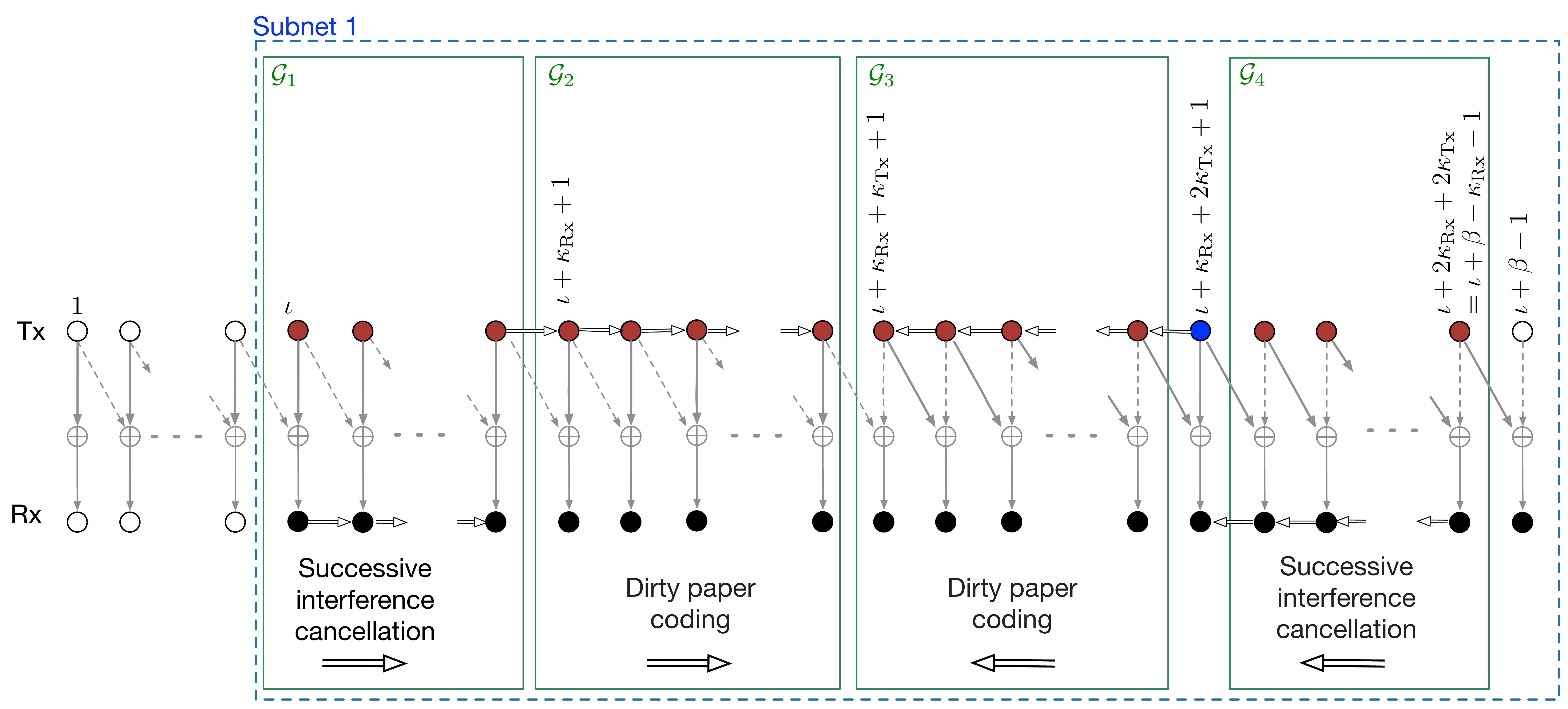}
\caption{An illustration of $\s{G}_1$, $\s{G}_2$, $\s{G}_3$ and $\s{G}_4$ in Subnet 1. The transmitter corresponding to the special source message $M_{\i +  \kappa_\Rx+2 \kappaT+1} $ is coloured in blue. The figure also shows the Tx- and Rx-conferencing links that are effectively used in our scheme, and it indicates that the desired communication paths for group~1 and 2 messages are the direct links, and for group 3 and 4 messages the diagonal links.}
\label{Fig:Subnet1}
\end{center}
\end{figure}

Since the subnets are identical and do not interfere with one another, we need only describe the coding scheme for subnet $1$ (transmitters $\i, \ldots, \i +\beta-2$ and receives $\i, \ldots, \i+\beta-1$), where we need to communicate  source messages
\begin{equation}\label{Eqn:Group1Messages}
\big( M_{\i}\ M_{\i+1}\ \ldots\ M_{ \i +\beta-1} \big).
\end{equation}
Source message $M_{ \i +\beta-\kappaR-1}$ will be handled in a special way: We will split it into independent sub-messages
\begin{equation}\label{Eqn:ExtraMessage}
M_{ \i +\beta-\kappaR-1} = \big(M_{ \i +\beta-\kappaR-1}^{\text{Tx}} \
M_{ \i +\beta-\kappaR-1}^{\text{Rx}}\big),
\end{equation}
where $M_{ \i +\beta-\kappaR-1}^{\text{Tx}}$ has rate $R_{ \i +\beta-\kappaR-1}^{\text{Tx}}$ and  $M_{ \i +\beta-\kappaR-1}^{\text{Rx}}$ has rate $R_{ \i +\beta-\kappaR-1}^{\text{Rx}}$. We will partition the remaining source messages in~\eqref{Eqn:Group1Messages} into four groups $\mathcal{G}_1, \s{G}_2, \s{G}_3,$ and $\mathcal{G}_4$, see Figure~\ref{Fig:Subnet1}. In our scheme, the messages in each group are transmitted using a different strategy. The special source message $M_{ \i +\beta-\kappaR-1}^{\text{Tx}}$ is transmitted in the same way as source messages in $\mathcal{G}_3$, and  source message $M_{ \i +\beta-\kappaR-1}^{\text{Rx}}$ is transmitted in the same way as source messages in $\mathcal{G}_4$. 

We next describe how to partition the messages into groups $\s{G}_1,\ldots, \s{G}_4$ and sketch how to communicate the source messages in these groups. The communications are  described with more technical details in Appendix~\ref{app3}, where we also present an analysis.
\begin{itemize}
\item \emph{Group 1 (successive interference cancellation from left to right, see Figure~\ref{Fig:GroupG1}):} The first group of source messages,
\begin{equation}\label{eq:rxconfL}
\big\{M_{k} : k \in \s{G}_1 \big\}
\end{equation}
with
\begin{equation*}
\s{G}_1 := \big\{\i, \i+1, \ldots \i +\kappa_\Rx\big\},
\end{equation*}
is communicated using \emph{point-to-point channel codes} with \emph{successive interference cancellation} from left-to-right at the receivers. 
Specifically, each Tx~$k\in\s{G}_1$ uses a Gaussian point-to-point code of power $P$ to transmit its source message~$M_k$. Inputs $X_k^n(M_k)$ thus only depend on~$M_k$. 

The decoding procedure is depicted in Figure~\ref{Fig:GroupG1} and described with more technical details in Appendix~\ref{sec:Group1}.

Recall that we deactivated Tx~$\i-1$ (the last transmitter in the previous subnet).  Rx~$\i$  (the first receiver in group 1) thus observes channel outputs
\begin{equation}\label{eq:channeli}
Y_\i^n = X_\i^n(M_\i)+Z_\i^n,
\end{equation}
based on which it decodes its desired source message $M_\i$. 

Rx~$\i$ also describes its decoded source message
$\hat{M}_\i$ over the conferencing link to Rx~$\i+1$ (its immediate right-neighbour). 

Rx~$\i+1$ uses this conferencing message to reconstruct $\alpha_{\i+1} X_{\i}^n(\hat{M}_\i)$. 
It then forms 
\begin{equation*}
\hat{Y}_{\i+1}^n={Y}_{\i+1}^n-\alpha_{\i+1} X_{\i}^n(\hat{M}_\i)
\end{equation*}
and decodes source message $M_{\i+1}$ based on this difference. Rx~$\i+1$ also describes $\hat{M}_{\i+1}$ over the conferencing link to Rx~$\i+2$ (its immediate right-neighbour).

Notice that whenever $\hat{M}_\i=M_\i$, Rx~$\i+1$ decodes $M_{\i+1}$ based on the interference-free signal
\begin{equation}\label{eq:channeli2}
\hat{Y}_{\i+1}^n= X_{\i+1}^n(M_{\i+1}) + Z_{\i+1}^n.
\end{equation}

The same procedure is continued for receivers~$k=\in \i+2,\ldots, \i+\kappaR$. Specifically, each of these receivers performs the following three steps:
\begin{enumerate}
\item It  reconstructs interference $\alpha_k X_{k-1}^n(\hat{M}_{k-1})$ using the conferencing message $\hat{M}_{k-1}$ obtained from Rx~$k-1$. 
\item It forms  the presumingly interference-free signal
\begin{equation*}
\hat{Y}_{k}^n=Y_k^n-\alpha_{k} \hat{X}_{k-1}^n(\hat{M}_{k-1}),
\end{equation*} and it decodes source message $M_k$  based on this difference.
\item  It sends the decoded source message $\hat{M}_k$ over the conferencing link to Rx~$k+1$. 

The last receiver~$\i+\kappaR$ does not send anything over the conferencing link; it skips this third step.
\end{enumerate}

Notice that the described scheme requires only left-to-right Rx-conferencing; no Tx-conferencing and no right-to-left conferencing. Also, Rx~$\i+j-1$ (the $j$-th receiver of group~$\s{G}_1$), for $j\in\{1,\ldots, \kappaR-1\}$, has to wait until it obtains the Rx-conferencing message from its left-neighbour before it can start performing above three steps. It can thus send its own conferencing message~$\hat{M}_{\i+j-1}$ only in Rx-conferencing round~$j$. 

Finally, we notice that for each $k$, if Rx~$k-1$ has correctly decoded its message, i.e., $\hat{M}_{k-1}={M}_{k-1}$, then Rx~$k$ decodes the source message $M_k$ based on the interference-free signal $\hat{Y}_{k}^n=X_k^n+Z_k^n$. 
Source messages $M_\i,\ldots, M_{\i+\kappaR}$ can thus be decoded with vanishingly small probability of error as $n\to \infty$, whenever (see also Lemma~\ref{Lem:Group1} in Appendix~\ref{sec:Group1})
\begin{equation}\label{eq:R_G1}
R_k < \frac{1}{2} \log (1+P), \qquad \forall k\in\s{G}_1
\end{equation}
and
\begin{equation}
R_\Rx>R_k,\qquad \forall k\in\s{G}_1.
\end{equation}
\begin{figure}[h!]
\begin{center}
\includegraphics[width=0.7\textwidth]{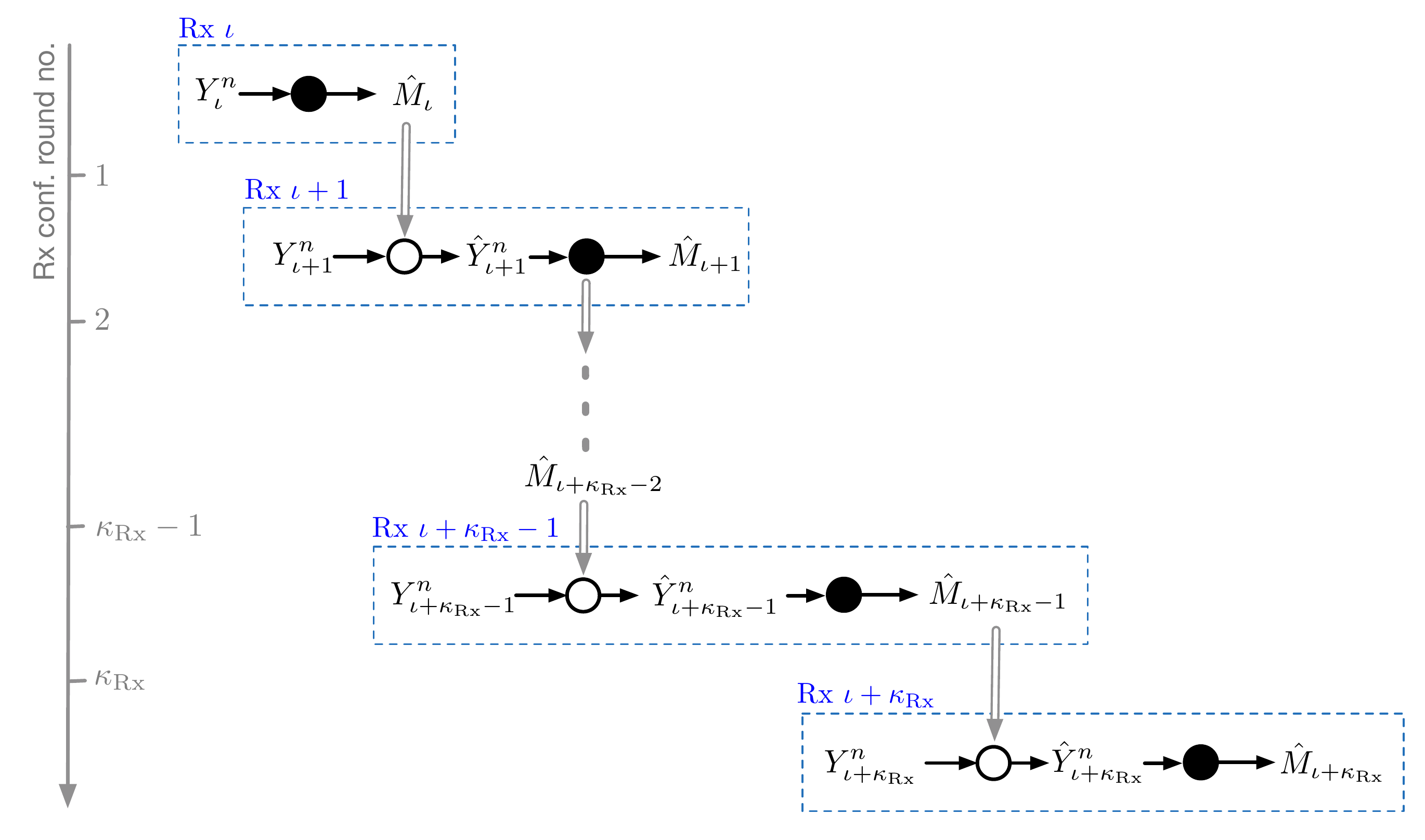}
\caption{Group 1: Left-to-right successive cancellation at the receivers. White circles represent interference cancellation 
and the black circles represent channel decoding. Top-down arrows represent communications over left-to-right Tx-conferencing links.}
\label{Fig:GroupG1}
\end{center}
\end{figure}

\item \emph{Group 2 (dirty paper coding from left to right, see Figure~\ref{Fig:LeftDirtyPaperCoding}):}
The second group of source messages,
\begin{equation}\label{eq:txconfL}
\big\{M_{k}\ : k \in \s{G}_2 \big\}
\end{equation}
with
\begin{equation*}
\s{G}_2 := \{ \i + \kappaR + 1, 
\i + \kappaR + 2,
\ldots,
\i + \kappa_\Rx+\kappaT
\big\},
\end{equation*}
will be communicated using \emph{dirty paper coding} to mitigate the interference from the left. The encoding procedure is depicted in Figure~\ref{Fig:LeftDirtyPaperCoding} and is explained with more technical details in~Appendix~\ref{sec:Group2}.
 \begin{figure}
 \begin{center}
 \includegraphics[width=0.8\textwidth]{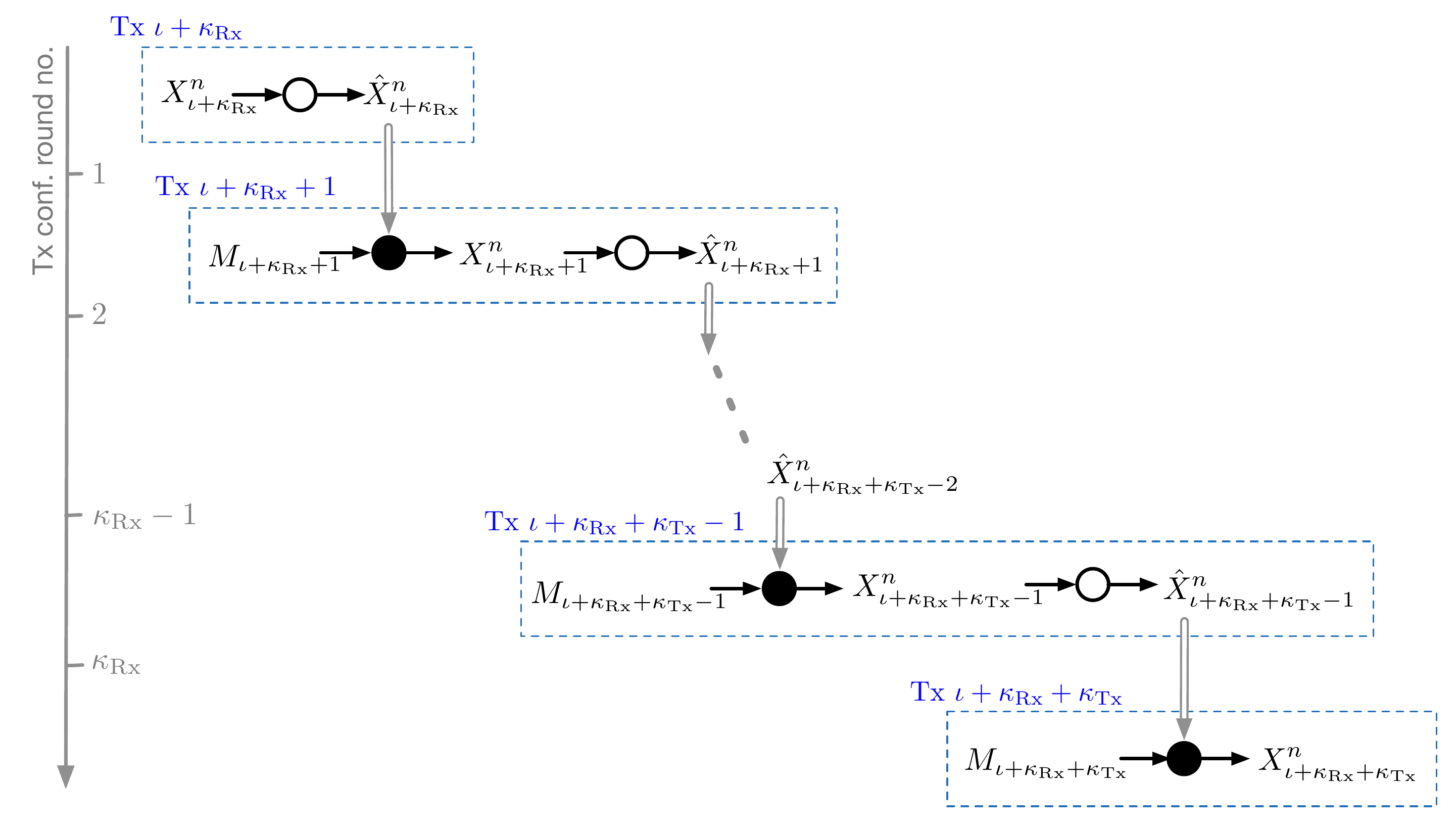}
 \caption{Group 2: Tx dirty paper coding from left to right. The dark circles represent \emph{dirty paper channel} encoders, and the white circles represent \emph{vector Gaussian} quantisers. Top-down arrows represent communications of quantised input signals over the Tx-conferencing links to right-neighbours.}
 \label{Fig:LeftDirtyPaperCoding}
 \end{center}
 \end{figure}

To facilitate dirty-paper coding at Tx~$\i+\kappaR+1$ (the first transmitter in group~2), Tx~$\i+\kappaR$ (the last transmitter in group 1) quantises its inputs signal $X_{\i+\kappaR}^n$ using a rate-$\frac{1}{2} \log (1+P)$ quantiser. It sends the quantisation message to this Tx~$\i+\kappaR+1$. 

 Tx~$\i+\kappaR+1$ reconstructs the quantised inputs $\hat{X}_{\i+\kappaR}^n$ and encodes its source message $M_k$ using a power~$P$ dirty-paper code that eliminates  interference $\alpha_{\i+\kappaR+1}\hat{X}_{\i+\kappaR}^n$. 
 
 Tx~$\i+\kappaR+1$ also quantises its produced inputs $X_{\i+\kappaR+1}^n$ using a rate-$\frac{1}{2} \log (1+P)$ quantiser, and sends the quantisation message to Tx~$\i+\kappaR+2$ (its right-neighbour).
 
 Rx~$\i+\kappaR+1$  decodes source message $M_{\i+\kappaR+1}$ by
applying dirty-paper decoding to its outputs 
 \begin{equation*}
 Y_{\i+\kappaR+1}^n=\alpha_{\i+\kappaR+1}{X}_{\i+\kappaR}^n+{X}^n_{\i+\kappaR+1}+{Z}^n_{\i+\kappaR+1}.
 \end{equation*} 
 Notice that $\alpha_{\i+\kappaR+1}\hat{X}_{\i+\kappaR}^n-\alpha_{\i+\kappaR+1}{X}_{\i+\kappaR}^n$ has variance close to $n\alpha_{\i+\kappaR+1}^2 \frac{P}{P+1}$ and thus for $P\gg 1$ the dirty-paper code precancels the predominant part of the interfering signal $\alpha_{\i+\kappaR+1}{X}_{\i+\kappaR}^n$
 \\
 
 The procedure is repeated for transmitters and receivers $k=\i+\kappaR+2, \ldots, \i+\kappaR+\kappaT$. Each such Tx~$k$ performs the following three steps:
\begin{enumerate}
\item Using the conferencing message from Tx~$k-1$, it reconstructs the quantised signal $\hat{X}_{k-1}^n$. 

\item It encodes and transmits its source message  $M_k$ using a power-$P$ dirty-paper code that eliminates  interference $\alpha_k \hat{X}_{k-1}^n$.  
\item It quantises its input signals $X_k^n$ with a rate-$\frac{1}{2} \log (1+P)$ quantiser 
and sends the produced quantisation message over the conferencing link  to Tx~$k+1$. 


 Tx~$\i + \kappa_\Rx+\kappaT$ (the last transmitter in group $\s{G}_2$)  sends no conferencing message; it skips step 3.
\end{enumerate} 

Each corresponding Rx~$k$ decodes its desired source message $M_k$ using dirty-paper decoding based on its outputs $Y_k^n$. \\

Notice that the described scheme requires only left-to-right Tx-conferencing; no right-to-left Tx-conferencing and no Rx-conferencing. Also, Tx~$\i+\kappaR+j$, (the $j$-th transmitter of group $\s{G}_2$), for $j\in\{1,\ldots,\kappaT-1\}$, has to wait until it obtains the Tx-conferencing message from its left-neighbour before it can perform above three steps. It can thus send its own 
Tx-conferencing message to its right-neighbour only in Tx-conferencing round $j+1$. 

Finally, as we show in detail in Appendix~\ref{sec:Group2}, through a careful design of the vector-quantisers and the dirty-paper codes and when $P\gg1$,  the predominant part  of the interference $\alpha_k X_{k-1}^n$ experienced at Rx~$k$ can be precanceled. As a consequence, source messages $M_{\i+\kappaR+1},\ldots, M_{\i+\kappaR+\kappaT}$ can  be decoded with vanishingly small probability of error as $n\to \infty$, whenever (see also Lemma~\ref{Lem:Group2} in Appendix~\ref{sec:Group2})
\begin{equation}\label{eq:R_G2}
R_k < \frac{1}{2} \log \left( 1+ \frac{P}{1+ \alpha_k^2 \frac{P}{P+1}} \right), \qquad \forall k\in\s{G}_2
\end{equation}
and
\begin{equation}
R_{\Tx} > \frac{1}{2} \log( 1+{P}).
\end{equation}

\end{itemize} 
\bigskip
Source messages in groups 1 and 2 were transmitted over the ``direct" links from a Tx~$k$ to its corresponding Rx~$k$. The ``diagonal" links from a Tx~$k$ to its right-neighbouring Rx~$k+1$ only carried interference that had to be mitigated. 
In contrast, source messages in groups 3 and 4 are transmitted over  the ``diagonal" links,  whereas the ``direct" links carry interference that has to be mitigated. 
Without conferencing, there exists no desired communication path over the diagonal links, since the source message desired by Rx~$k$ is \emph{a priori} unknown at its left-neighbour Tx~$k-1$. Setting up such a path requires concatenating the diagonal link from Tx~$k-1$ to Rx~$k$ with  a preceding  right-to-left Tx-conferencing from Tx~$k$ to Tx~$k-1$ or a  subsequent right-to-left Rx-conferencing from Rx~$k$ to Rx~$k-1$, see \eqref{eq:diagonal_path1} and \eqref{eq:diagonal_path2} ahead.

We now describe transmission of messages in group 4, follows by transmission of messages in group 3.

\begin{itemize}
\item \emph{Group 4 (successive interference cancellation from right to left):} The fourth group of messages,
\begin{equation}\label{eq:rxconfR}
\big\{ M^\Rx_{\i + \beta-\kappaR-1} \big\} 
\cup
\big\{M_k : k \in \s{G}_4\big\}
\end{equation}
with
\begin{equation*}
\s{G}_4 := 
\big\{\i + \beta-\kappaR,\ldots,\i +\beta-2 \},
\end{equation*} 
is communicated using \emph{point-to-point channel codes} with right-to-left \emph{successive interference cancellation} at the decoders.  (See Appendix~\ref{sec:Group4} for more technical details.)

Specifically, each source message $M_k$\footnote{For ease of notation, in the following paragraph we write  simply $M_{\i + \beta-\kappaR-1}$ for $M_{\i + \beta-\kappaR-1}^{\textnormal{Rx}}$} is sent over the following communication  path:
\begin{equation}\label{eq:diagonal_path2}
\textnormal{Tx}~k\; \longrightarrow\; \textnormal{Rx}~{k+1}\; \longrightarrow\; \textnormal{Rx}~k.
\end{equation}
Each Tx~$k\in\s{G}_4$ uses a Gaussian point-to-point code of power $P$ to transmit its source message~$M_k$. 

Rx~$\i+\beta-1$ (the last receiver in the subnet) observes channel outputs
\begin{equation}\label{eq:channelit}
Y_{\i+\beta-1}^n = \alpha_{\i+\beta-1}X_{\i+\beta-2}^n(M_{\i+\beta-2})+Z_{\i+\beta-1}^n,
\end{equation}
because we silenced its corresponding Tx~${\i+\beta-1}$. It  decodes  source message $M_{{\i+\beta-2}}$ from these channel outputs and describes the decoded message
$\hat{M}_{\i+\beta-2}$ 
 over the conferencing link to Rx~${\i+\beta-2}$ (its immediate left-neighbour). 

Rx~${\i+\beta-2}$ declares the obtained conferencing message $\hat{M}_{\i+\beta-2}$  as its guess of  ${M}_{\i+\beta-2}$. 
It further uses the conferencing message to reconstruct $ X_{\i+\beta-2}^n(\hat{M}_{\i+\beta-2})$ and forms 
\begin{equation*}
\hat{Y}_{\i+\beta-2}^n={Y}_{\i+\beta-2}^n- X_{\i+\beta-2}^n(\hat{M}_{\i+\beta-2}).
\end{equation*}
Rx~$\i+\beta-2$ finally decodes  source message $M_{\i+\beta-3}$ from this difference and describes $\hat{M}_{\i+\beta-3}$ over the conferencing link to Rx~$\i+\beta-3$.

Whenever $\hat{M}_{\i+\beta-2}={M}_{\i+\beta-2}$,
\begin{equation}\label{eq:channeli2}
\hat{Y}_{\i+\beta-2}^n= \alpha_{\i+\beta-2}X_{\i+\beta-3}^n(M_{\i+\beta-3}) + Z_{\i+\beta-2}^n,
\end{equation}
and Rx~$\i+\beta-2$ decodes source message $M_{\i+\beta-3}$ based on an interference-free signal.\\

The described procedure is repeated for receivers~$k=\i+\beta-3, \ldots, \i+\beta-\kappaR-1$ in decreasing order. Specifically, each such 
Rx~$k$ performs four steps:
\begin{enumerate}
\item  Using the conferencing message $\hat{M}_{k}$ from Rx~$k+1$, it reconstructs the ``interference"~$X_{k}^n(\hat{M}_k)$.
\item It forms $\hat{Y}_k^n:=Y_k^n-{X}_k^n(\hat{M}_k)$  and decodes source message $M_{k-1}$ based on this difference.
\item  It sends the decoded source message $\hat{M}_{k-1}$ over the conferencing link to  Rx~$k-1$. 
\item It declares  $\hat{M}_{k}$ as its guess of source message $M_k$.
\end{enumerate}

Notice that the described scheme requires only right-to-left Rx-conferencing; no left-to-right Rx-conferencing nor Tx-conferencing.  Rx~$\i+\beta-1$ (the last receiver in the subnet) sends its conferencing message $\hat{M}_{\i+\beta-2}$ in the first Rx-conferencing round. For each $j\in\{2, \ldots, \kappaR\}$,  Rx~$\i+\beta-j$ (the $j$-th right-most receiver in the subnet) has to wait until it obtains the conferencing message from its right-neighbour before it can start performing above four steps. It can thus send its own conferencing message $\hat{M}_{\i+\beta-j-1}$ only in Rx-conferencing round~$j$.

Finally, we notice that if Rx~$k+2$'s decoding was successful, i.e., $\hat{M}_{k+1}={M}_{k+1}$, then $\hat{X}_{k+1}^n=X_{k+1}^n$ and $\hat{Y}_{k+1}^n=\alpha_{k+1}X_{k}^n+Z_{k+1}^n$, and Rx~$k+1$ can thus decode source message $M_{k}$ based on an interference-free signal. \\

Consequently,  source messages $M_{\i+\beta-\kappaR-1}^{\Rx}, M_{\i+\beta-\kappaR}, \ldots, M_{\i+\beta-2}$  can be decoded with vanishingly small probability of error as $n\to \infty$, whenever
\begin{subequations}\label{eq:R_G4}
\begin{equation}
R_k < \frac{1}{2} \log (1+\alpha_{k+1}^2P), \qquad k\in\s{G}_4,
\end{equation}
 \begin{equation}\label{eq:prelog17a}
R_{\i+\beta-\kappaR-1}^{\Rx} <  \frac{1}{2} \log (1+\alpha_{\i+\beta-\kappaR}^2P),
 \end{equation}
\end{subequations}
 and 
\begin{equation}
R_\Rx > \max\Big\{R_{\i+\beta-\kappaR-1}^{\Rx}, \ R_{\i+\beta-\kappaR},\  \ldots, R_{\i+\beta-2}\Big\}.
\end{equation}

\item \emph{Group 3 (dirty paper coding from right to left):} The third group of messages 
\begin{equation}\label{eq:txconfR}
\big\{ M_k : k \in \s{G}_3\big\} 
\cup 
\big\{ M^\Tx_{\i + \beta-\kappaR-1} \big\}
\end{equation}
with
\begin{equation*}
\s{G}_3 := \big\{\i +\beta-\kappaR-\kappaT,
\ldots, \i + \beta-\kappaR-2 \big\},
\end{equation*}
uses dirty paper coding to mitigate the interference from the right. The desired communication path of a message $M_k$ is 
\begin{equation}\label{eq:diagonal_path1}
\textnormal{Tx}~k\; \longrightarrow\; \textnormal{Tx}~k-1\; \longrightarrow\; \textnormal{Rx}~k 
\end{equation}and thus involves Tx-conferencing from right-to-left.  More specifically, Tx~$k$ prepares a transmit signal $X_{k-1}^n$ that encodes message $M_k$. It then describes it  over the conferencing link to Tx~$k-1$, which will transmit this signal.\\

For simplicity, we temporarily fix $k=\i+\beta-\kappaR-1$. In the previous subsection we described how Tx~$k$ (the special transmitter of Figure~\ref{Fig:Subnet1}), generated its input signal $X_{k}^n$ in function of $M_{k}^{\textnormal{Rx}}$. Tx~$k$ now also encodes its second message $M_{k}^{\textnormal{Tx}}$ using a dirty-paper code of power $\alpha_{k}^2 \frac{P^2}{P+1}$ that mitigates its own input signal $X_{k}^n$. Denote the produced dirty-paper sequence by $\Psi_{\textnormal{DPC},k}\big(M_{k}^{\textnormal{Rx}}\big)$.

Tx~$k$ quantises a scaled version of the dirty-paper sequence, 
\begin{equation*}
\Phi_{k}^n:=\frac{P+1}{P}\Psi_{\textnormal{DPC},k}\big(M_{k}^{\textnormal{Rx}}\big)
\end{equation*} using a rate-$\frac{1}{2}\log(1+P)$ quantiser and sends the resulting quantisation message over the conferencing link to Tx~$k-1$ (its left neighbour). 

Tx~$k-1$ reconstructs the quantised sequence $\hat{\Phi}_{k}^n$ and transmits 
\begin{equation*}
X_{k-1}^n= \alpha_{k}^{-1} \hat{\Phi}_k^n
\end{equation*}
over the interference network. 

Rx~$k$ decodes source message $M_{k}^{\textnormal{Tx}}$ by applying dirty-paper decoding to its outputs $Y_{k}^n$. \\

Notice that if there was no quantisation error,
\begin{equation}\label{eq:ass}
\Delta_k^n:=\hat{\Phi}_k^n- \Phi_{k}^n=0,
\end{equation}
then 
Rx~$k$ would observe outputs
\begin{equation*}
Y_k^n=\Psi_{\textnormal{DPC},k}\big(M_{k}^{\textnormal{Rx}}\big) + X_{k^n} + Z_k^n.
\end{equation*}
In this case, since the applied  dirty-paper code precancels the ``interference" $X_k^n$, source message $M_k$ could be transmitted with the same rates as over an interference-free channel.

Now, zero quantization error \eqref{eq:ass} is very unlikely. However,  when the quantiser is chosen as in Appendix~\ref{sec:Group3}, the normalized variance  $\textnormal{Var}\big[\frac{1}{n}\Delta_k^n\big]$ approaches $\alpha_k^2\frac{1}{P+1}$ and is bounded in  $P$. Treating the quantisation $\Delta_k^n$ simply as an additional noise, will thus not degrade the prelog rate of the source message $M_k$. (See Appendix~\ref{sec:Group3} for a more detailed analysis.)\\

Let now $k=\i+\beta-\kappaR-2$, one less than before. 

Tx~$k$ (which before was Tx~$k-1$) also prepares the transmit signal of its left-neighbour. Specifically, Tx~$k$ encodes its own source message $M_{k}$ using a dirty-paper code of power $\alpha_{k}^2 \frac{P^2}{P+1}$ that mitigates its own input signal $X_k^n$. Let $\Psi_{\textnormal{DPC},k}^n(M_{k})$ denote the produced dirty-paper sequence. Tx~$k$ applies a rate-$\frac{1}{2}\log(1+P)$ quantiser to the scaled sequence 
\begin{equation*}
\Phi_{k}^n:= \frac{P+1}{P}\Psi_{\textnormal{DPC},{k}}^n(M_{k})
\end{equation*} and sends the resulting quantisation message to Tx~$k-1$.\\

This process is repeated for transmitters and receivers $k=\i+\beta-\kappaR-3, \ldots, \i+\beta-\kappaR-\kappaT$ in decreasing order. Each such Tx~$k$ performs the following four steps:\footnote{See Appendix~\ref{sec:Group3} for more technical details.} 
\begin{enumerate}
\item Using the conferencing message from Tx~$k+1$, it reconstructs the quantised signal $\hat{\Phi}_{k+1}^n$. 
\item It  transmits $X_k^n= \alpha_{k+1}^{-1}\hat{\Phi}_{k+1}^n$ over the network.
\item  It encodes source message $M_k$ using dirty-paper coding of power $\alpha_k^2\frac{P^2}{P+1}$ that mitigates \emph{its own inputs $X_k^n$}. It then  forms ${\Phi}_{k}^n= \frac{P+1}{P}\Psi_{\textnormal{DPC},k}^n(M_k)$, where $\Psi_{\textnormal{DPC},k}^n(M_k)$ denotes the produced dirty-paper sequence.
\item It quantises ${\Phi}_{k}^n$ using a rate-$\frac{1}{2}\log(1+P)$ quantiser and sends the resulting quantisation bits over the conferencing link to Tx~$k-1$.
\end{enumerate}  
Tx~$\i+\kappaR+\kappaT+1$ (the left-most transmitter only performs steps 1) and 4), but prepares and sends no conferencing message. 

Each receiver applies dirty-paper decoding to decode its desired message $M_k$ based on its outputs $Y_k^n$. \\

Notice that the described scheme requires right-to-left Tx-conferencing; no left-to-right Tx-conferencing or Rx-conferencing are needed. Special Tx~$ \i + \kappaR +2\kappaT+1$ sends its conferencing message in Tx-conferencing round $1$. Each Tx~$\i+\beta-\kappaR-\j$, for $j\in\{2,\ldots, \kappaT-1\}$ can send its conferencing message only after receiving the conferencing message from its right-neighbour. Tx~$\i+\beta-\kappaR-\j$ can thus send its conferencing message only in Tx-conferencing round $j+1$. 

Finally, as we show in detail in Appendix~\ref{sec:Group3}, through a careful design of the quantisers  and the dirty-paper codes, source messages $M_{\i+\beta-\kappaR-\kappaT}, \ldots, M_{\i+\beta-\kappaR-1}^{\textnormal{Tx}}$ can be decoded with vanishingly small probability of error as $n \to \infty$, whenever
\begin{subequations}\label{eq:R_G3}
\begin{IEEEeqnarray}{rCl}
R_k < \frac{1}{2} \log\left( 1+ \frac{\alpha_k^2 \frac{P^2}{P+1}}{1+\alpha_k^2\frac{P}{P+1}}\right), \hspace{1cm} k\in\mathcal{G}_3, 
\end{IEEEeqnarray}
 \begin{equation}\label{eq:prelog17a}
R_{\i+\beta-\kappaR-1}^{\Tx} < \frac{1}{2} \log\left( 1+ \frac{ \frac{P^2}{P+1} \alpha_{\i+\beta-\kappaR-1}^2}{1+ \frac{P}{P+1}\alpha_{\i+\beta-\kappaR-1}^2}\right)
 \end{equation}
\end{subequations}
and
\begin{equation}
R_{\Tx} > \frac{1}{2} \log( 1+{P}).
\end{equation}
\end{itemize}



\subsection{Tx-Conferencing in Period~2} \label{sec:tx_conferencing}
If in the previous subsection~\ref{sec:tx_rx_conferencing} we set everywhere $\kappa_\Rx=0$, and hence $\mathcal{G}_1=\mathcal{G}_4=\emptyset$, we obtain our scheme for period 2 in case of Tx-conferencing only. In this case, there is no need for a message ${M}_{\i+\kappaT+2 \kappaR+1}^{\textnormal{Rx}}$ and we can simply set ${M}_{\i+\kappaT+2 \kappaR+1}^{\textnormal{Tx}}={M}_{\i+\kappaT+2 \kappaR+1}$.


\subsection{No Conferencing in Period 3}  \label{sec:no_conferencing}

We silence all odd transmitters. This splits the network into a set of $\lceil \frac{K}{2}\rceil$ 
parallel Gaussian point-to-point channels. We use optimal point-to-point codes over these channels. 

\subsection{Analysis}

\subsubsection{Analysis of Period~1}

As argued in \eqref{eq:R_G1}, \eqref{eq:R_G2},  \eqref{eq:R_G4} and \eqref{eq:R_G3} (see also lemmas~\ref{Lem:Group1}--\ref{Lem:Group3}), in each subnet we can transmit $2 \kappaT+2\kappaT+1$ (sub)messages each of MG~1. 
Thus, over the entire network which consists of $\gamma$ subnets, our scheme achieves a MG of
 \begin{equation}\label{eq:mugain}
\gamma (2 \kappaT+2 \kappa_\Rx+1),
 \end{equation}
 and a per-user MG of
 \begin{equation}\label{eq:S_P1}
\S_{\textnormal{period1}}= \frac{ 2 \kappaT+2 \kappa_\Rx+1}{ 2 \kappaT+2 \kappa_\Rx+2}.
 \end{equation} 
 
 We now analyse the communication over the conferencing links. We have 
for each subnet $g=0,\ldots, \gamma-1$ (see lemmas~\ref{Lem:Group1}--\ref{Lem:Group3}),
\begin{itemize} 
\item Only the $\kappaT$ consecutive transmitters~$\i+ g \beta+ \kappaR, \ldots, \i+g\beta+\kappaR+\kappaT-1$  send conferencing messages to their right-neighbours. Each of these messages corresponds to a decoded source message. Since we transmit all source messages at rates below $\frac{1}{2}\log(1+P)$, also the rates of the conferencing messages do not exceed $\frac{1}{2}\log(1+P)$.
\item  Only the $\kappaT$ consecutive transmitters~$\i+(g+1)\beta-\kappaR-\kappaT, \ldots, \i+(g+1)\beta-\kappaR-1 $ send conferencing messages to their left-neighbours. Each of these messages corresponds to a decoded source message. Since we transmit all source messages at rates below $\frac{1}{2}\log(1+P)$, also the rates of the conferencing messages do not exceed $\frac{1}{2}\log(1+P)$.
\item  Only the $\kappaR$ consecutive receivers~$\i+ g \beta, \ldots, \i+g\beta+\kappaR-1$ send  conferencing messages to their right-neighbours. Each of these messages corresponds to a rate-$\frac{1}{2}\log(1+P)$ quantisation message.\footnote{More precisely, the rate should be slightly larger than  $\frac{1}{2}\log(1+P)$. Through standard continuity considerations one can show that this does not change the set of achievable rates. It is thus a minor technicality, which we ignore.}
\item  Only the $\kappaR$ consecutive  receivers~$\i+ (g+1) \beta-\kappaR, \ldots, \i+(g+1)\beta-1$ send  conferencing messages to their left-neighbours. Each of these messages corresponds to a rate-$\frac{1}{2}\log(1+P)$ quantisation message.
\end{itemize}

We conclude that each Tx-conferencing link is used a $\frac{\kappaT}{2\kappaT+2\kappaR+2}$-th of the time 
and at a total rate not exceeding  
\begin{equation}\label{eq:Tx_P1}
\mu_{\Tx,\textnormal{Period1}}=\frac{\kappaT}{2\kappaT+2\kappaR+2}\cdot \frac{1}{2}\log(1+P).
\end{equation}
Similarly, each Rx-conferencing link is used a $\frac{\kappaR}{2\kappaT+2\kappaR+2}$-th of the time and at a total rate not exceeding
\begin{equation}\label{eq:Rx_P1}
\mu_{\Rx,\textnormal{Period1}}=\frac{\kappaR}{2\kappaT+2\kappaR+2}\cdot \frac{1}{2}\log(1+P).
\end{equation}


\subsubsection{Analysis of Period~2}
Specializing the findings of the previous section to $\kappaR=0$, the per-user~MG achieved in period~2 is
\begin{equation}\label{eq:S_P2}
\S_{\textnormal{period2}}=\frac{2 \kappaT+1}{2 \kappaT+2}.
\end{equation}
The tx-conferencing links are used at rate
\begin{equation}\label{eq:Tx_P2}
\mu_{\Tx,\textnormal{Period2}}=\frac{\kappaT}{2\kappaT+2}\cdot \frac{1}{2}\log(1+P),
\end{equation}
and 
the rx-conferencing links are not used at all, 
\begin{equation}\label{eq:Rx_P2}
\mu_{\Rx,\textnormal{Period2}}=0.
\end{equation}
%

\subsubsection{Analysis of Period~3} 
During the third period the achieved per-user MG is
\begin{equation}\label{eq:S_P3}
\S_{\textnormal{period3}}= \frac{1}{2}.
\end{equation}
The conferencing links are not used at all,
\begin{IEEEeqnarray}{rCl}\label{eq:Tx_P3}
\mu_{\Tx,\textnormal{Period3}} &=&0 \\
\mu_{\Rx,\textnormal{Period3}} &=&0.\label{eq:Rx_P3}
\end{IEEEeqnarray}

\subsubsection{Analysis of Overall Scheme} We first analyse the communication over the conferencing links. Under assumption~\eqref{eq:assumption}, there is tx-conferencing in periods~1 and 2. Given the length of the periods in
 \eqref{eq:choiceN},  and the tx-conferencing rates in periods 1 and 2, \eqref{eq:Tx_P1} and \eqref{eq:Tx_P2}, in total each tx-conferencing link is used at rate not exceeding  
 \begin{IEEEeqnarray}{rCl}\
 \lefteqn{
 \frac{N_1}{n} \mu_{\Tx,\textnormal{period1}} +  \frac{N_2}{n} \mu_{\Tx,\textnormal{period2}}} \qquad \nonumber\\
  & = & 
\left(\frac{N_1}{n} \cdot \frac{\kappaT}{2\kappaT+2\kappaR+2} + \frac{N_2}{n}\cdot \frac{\kappaT}{2\kappaT+2}\right) \cdot \frac{1}{2}\log(1+P)  \nonumber \\
&  \leq &\left( \frac{\muR}{\kappaR}(2\kappaT+2\kappaR+2) \frac{\kappaT}{2\kappaT+2\kappaR+2} +\Big(\frac{\muT}{\kappaT}- \frac{\muR}{\kappaR}\Big)(2\kappaT+2)  \frac{\kappaT}{2\kappaT+2} \right)\cdot \frac{1}{2}\log(1+P)  \nonumber \\
&  \leq&{\muT} \cdot \frac{1}{2}\log(1+P)=\RT.
\end{IEEEeqnarray}
Our overall scheme thus respects the tx-conferencing rate constraints in \eqref{eq:txconferencing_constraint}.

The rx-conferencing links are used only in period 1 at rate not exceeding \eqref{eq:Rx_P1}. Thus, considering the length of this period~1 in \eqref{eq:choiceN}, in total each receiver conferencing link is used at a rate not exceeding
\begin{IEEEeqnarray}{rCl}
 \frac{N_1}{n} \mu_{\Rx,\textnormal{period1}} 
& =  & 
\frac{N_1}{n} \cdot \frac{\kappaR}{2\kappaT+2\kappaR+2} \cdot \frac{1}{2}\log(1+P)  \nonumber \\
&\leq  & \frac{\muR}{\kappaR}(2\kappaT+2\kappaR+2)\cdot \frac{\kappaR}{2\kappaT+2\kappaR+2} \cdot \frac{1}{2}\log(1+P) \nonumber \\
& = & {\muR} \cdot \frac{1}{2}\log(1+P)=\RR.
\end{IEEEeqnarray}
Our overall scheme thus also respects the rx-conferencing rate constraints in \eqref{eq:rxconferencing_constraint}.

We now analyse the per-user MG achieved by our overall scheme. It is given by
\begin{equation}
\S=\frac{N_1}{n} \S_{\textnormal{period1}}+ \frac{N_2}{n} \S_{\textnormal{period2}}+\frac{N_3}{n} \S_{\textnormal{period3}}.
\end{equation}
To evaluate this expression, we distinguish three cases: 
\begin{enumerate} 
\item When 
\begin{equation}
1\leq \frac{\muR}{\kappaR}(2\kappaT+2\kappaR+2),\end{equation}
 then $N_1=n$ and periods 2 and 3 don't exist. In this case,
 \begin{equation}
 \S= \S_{\textnormal{period1}}= \frac{2 \kappaT+2\kappaR+1}{2\kappaT+2\kappaR+2}.
 \end{equation}

\item When $1>\frac{\muR}{\kappaR}(2\kappaT+2\kappaR+2) $ and 
\begin{IEEEeqnarray}{rCl}
1< \frac{\muR}{\kappaR}(2\kappaT+2\kappaR+2) +\Big(\frac{\muT}{\kappaT}- \frac{\muR}{\kappaR}\Big)(2\kappaT+2)= 2\muT + 2 \muR + 2\frac{\muT}{\kappaT}, 
\end{IEEEeqnarray}
then $N_1=n\frac{\muR}{\kappaR}(2\kappaT+2\kappaR+2)$, $N_2= n- N_1$ and $N_3=0$, i.e., period 3 does not exist. In this case,
\begin{IEEEeqnarray}{rCl}\S & = & 
\frac{N_1}{n} \S_{\textnormal{period1}}+ \Big(1- \frac{N_1}{n} \Big) \S_{\textnormal{period2}}  \nonumber \\
& = & \frac{\muR}{\kappaR}(2\kappaT+2\kappaR+2) \cdot \frac{2 \kappaT+2\kappaR+1}{2\kappaT+2\kappaR+2} + \left( 1- \frac{\muR}{\kappaR}(2\kappaT+2\kappaR+2) \right) \frac{2 \kappaT+1}{2\kappaT+2} \nonumber \\
& = & \frac{2 \kappaT+1}{2\kappaT+2}  +  \frac{\muR}{\kappaR}(2\kappaT+2\kappaR+2) \left( \frac{2 \kappaT+2\kappaR+1}{2\kappaT+2\kappaR+2}  -  \frac{2 \kappaT+1}{2\kappaT+2} \right) \nonumber \\
&= & \frac{ 2\kappaT+1+2\muR}{2\kappaT+2}.
\end{IEEEeqnarray}
\item When 
\begin{equation*}
1> 2\muT + 2 \muR + 2\frac{\muT}{\kappaT},
\end{equation*}
 then $N_1=n\frac{\muR}{\kappaR}(2\kappaT+2\kappaR+2)$, $N_2=n \Big(\frac{\muT}{\kappaT}- \frac{\muR}{\kappaR}\Big)(2\kappaT+2)$ and $N_3=n-N_1-N-2>0$. In this case, 
\begin{IEEEeqnarray}{rCl}\S & = &
\frac{N_1}{n} \S_{\textnormal{period1}}+  \frac{N_2}{n}\S_{\textnormal{period2}}+ \left(1 - \frac{N_1}{n} - \frac{N_2}{n}\right) \S_{\textnormal{period3}}  \nonumber \\
&& =\frac{\muR}{\kappaR}(2\kappaT+2\kappaR+2) \cdot \frac{2 \kappaT+2\kappaR+1}{2\kappaT+2\kappaR+2} + \Big(\frac{\muT}{\kappaT}- \frac{\muR}{\kappaR}\Big)(2\kappaT+2)\cdot  \frac{2 \kappaT+1}{2\kappaT+2} \nonumber \\
& & + \left( 1- \frac{\muR}{\kappaR}(2\kappaT+2\kappaR+2) - \Big(\frac{\muT}{\kappaT}- \frac{\muR}{\kappaR}\Big)(2\kappaT+2)\right) \cdot \frac{1}{2} \nonumber\\
& = & \frac{ 1+ 2\muT+2\muR}{2}.
\end{IEEEeqnarray}
\end{enumerate} 

Combining all these findings proves  Theorem~\ref{Thm:Ach}.

\section{Proof of Theorem~\ref{Thm:Converse}}\label{Sec:Proof:Thm:Converse}


Since with an unlimited number of conferencing rounds we cannot do worse than with $\bfkappa=(\kappa_\Tx,\kappa_\Rx)$ conferencing rounds, by Theorem~\ref{Thm:UnlimitedConf},
\begin{equation}\label{eq:conv2}
\S(\bfkappa,\bfmu) 
\leq \S_\infty(\bfmu) \leq 
\frac{2\muT + 2\muR + 1}{2}.
\end{equation}

We now prove that also 
\begin{equation}\label{eq:conv3}
 \S(\bfkappa,\bfmu) \leq \frac{2 \kappaT+2\kappaR+1}{2\kappaT+2\kappaR+2}
\end{equation}
holds.
 Let us suppose that a genie provides each transmitter the source messages of the $\kappaT$ transmitters on its left and right; that is, transmitter $k$ is given source messages
\begin{equation*}
(M_{k - \kappaT}\ \ldots\ M_{k-1})
\quad
\text{and} 
\quad
(M_{k+1}\ \ldots\ M_{k+\kappaT}).
\end{equation*}
Let us also suppose that the genie provides to each receiver the exact channel outputs at the $\kappaR$ receivers on its left and right; that is, receiver $k$ is given
\begin{equation*}
(Y^n_{k-\kappaR}\ \ldots\ Y^n_{k-1})
\quad
\text{and} 
\quad 
(Y^n_{k+1}\ \ldots\ Y^n_{k+\kappaR}).
\end{equation*}
Lapidoth~\emph{et al.}~\cite[Cor.~2]{lapidothlevyshamaiwigger14} established that the per-user MG of this \emph{genie-aided} problem is given by $(2\kappaT + 2\kappaR + 1)/(2\kappaT + 2\kappaR + 2)$. Since the per-user MG of the genie-aided problem cannot be smaller than $\S(\bfkappa,\bfmu)$ in the problem at hand, we obtain upper bound~\eqref{eq:conv3}.
\vspace{2mm}

We now establish the converse bound~\eqref{eq:convres2} for the case $\muT=0$. The other parameters $\bfkappa = (\kappaT,\kappaR)$ and $\muR$ are arbitrary. By Proposition~\ref{Prop:MG}, 
\begin{equation*}
\S(0,\kappaR,0,\muR) = \S(1,\kappaR,0,\muR) = \cdots.
\end{equation*}
The genie-aided arguments in the preceding paragraph remain valid for $\bfkappa' = (0,\kappaR)$ and $\bfmu' = (0,\muR)$, so
\begin{equation*}
\S(\kappaT,\kappaR,0,\muR) = \S(0,\kappaR,0,\muR) \leq \frac{2\kappaR+1}{2\kappaR+2},
\end{equation*}
which establishes \eqref{eq:convres2}.
\vspace{2mm}

The converse bound~\eqref{eq:convres3} for $\muR = 0$ can be proved in a similar way.

\section{Proof of Converse to Theorem~\ref{Thm:UnlimitedConf}}\label{sec:converse_unlimited}
We prove the converse bound in \eqref{eq:converse_infinity},
\begin{IEEEeqnarray*}{rCl}
\S_\infty(\bfmu)
& \leq&  \frac{1+2\mu_\Tx+2\mu_{\Rx}}{2}. 
\end{IEEEeqnarray*}

Assume for the moment that $K$ is even,
and let 
\begin{equation}
\alpha_{\max}\triangleq \max_{k=2,\ldots, K} |\alpha_k|.
\end{equation}
Define  $\mathcal{I}_{\even}$ and $\mathcal{I}_{\odd}$ to be the sets of indices in $\{1,\ldots, K\}$ that are even and odd respectively:
\begin{IEEEeqnarray}{rCl}
\mathcal{I}_{\even} & \triangleq &  \big\{ k \in\{1,\ldots, K\} \colon k \textnormal{ is even}\big\} \\
\mathcal{I}_{\odd} & \triangleq & \big\{ k \in\{1,\ldots, K\} \colon k \textnormal{ is odd}\big\}.
\end{IEEEeqnarray}

Define further
\begin{IEEEeqnarray}{rCl}
\mathbf M_{\even}&\eqdef&\big\{ M_k\colon k\in\mathcal{I}_{\even} \big\}\\
\mathbf M_{\odd}&\eqdef& \big\{ M_k\colon k\in\mathcal{I}_{\odd} \big\}\\
\mathbf X_{\even}^n&\eqdef& \big\{ X_k^n \colon k\in\mathcal{I}_{\even} \big\}\\
\mathbf Y_{\even}^n&\eqdef& \big\{ Y_k^n \colon k\in\mathcal{I}_{\even} \big\}\\
\mathbf Y_{\odd}^n&\eqdef& \big\{ Y_k^n \colon k\in\mathcal{I}_{\odd} \big\}
\end{IEEEeqnarray}
\begin{IEEEeqnarray}{rCl}
\mathbf G_{\noises,\even}^n& \eqdef&  \bigg(\alpha_2 Z_{1}^n-Z_2^n, \alpha_4Z_{3}^n-Z_4^n, \nonumber \\ 
& & \hspace{1cm} \alpha_6Z_{5}^n-Z_6^n, \ldots, \alpha_K Z_{K-1}^n-Z_K^n\bigg) \IEEEeqnarraynumspace
\end{IEEEeqnarray}
and
\begin{IEEEeqnarray}{rCl}
\mathbf G_{\noises,\odd}^n& \eqdef&  \bigg(\alpha_3Z_{2}^n-Z_3^n, \alpha_5Z_{4}^n-Z_5^n, \nonumber \\ 
& & \hspace{1cm} \alpha_7Z_{6}^n-Z_7^n, \ldots, \alpha_{K-1}Z_{K-2}^n-Z_{K-1}^n\bigg) \nonumber \\
\end{IEEEeqnarray}
Also, let $\mathbf U_{\even\to\odd}$ denote the set of all cooperation messages that are sent from transmitters with even indices to transmitters with odd indices: 
\begin{equation}
\mathbf U_{\even\to\odd} \triangleq \Big\{ U_{k'\to k}^{(1)}, \ldots,  U_{k'\to k}^{(\kappaT)} \colon k' \in \mathcal{I}_{\even}, \;k\in\mathcal{I}_{\odd}\Big\}.
\end{equation}
Similarly, let $\mathbf U_{\odd\to\even}$ denote the set of all cooperation messages that are sent from transmitters with odd indices to transmitters with even indices: 
\begin{equation}
\mathbf U_{\odd\to\even} \triangleq \Big\{ U_{k'\to k}^{(1)}, \ldots,  U_{k'\to k}^{(\kappaT)}\colon k' \in \mathcal{I}_{\odd}, \;k\in\mathcal{I}_{\even}\Big\}.
\end{equation}
 In the same way, let  $\mathbf V_{\even\to\odd}$ denote the set of all cooperation messages that are sent from receivers with even indices to receivers with odd indices, 
 \begin{equation}
\mathbf V_{\even\to\odd} \triangleq \Big\{ V_{k'\to k}^{(1)}, \ldots,  V_{k'\to k}^{(\kappaR)} \colon k' \in \mathcal{I}_{\even}, \;k\in\mathcal{I}_{\odd}\Big\}.
\end{equation}
 and let $\mathbf V_{\odd\to\even}$ denote the set of all cooperation messages that are sent from receivers with odd indices to receivers with even indices:
 \begin{equation}
\mathbf V_{\odd\to\even} \triangleq \Big\{ V_{k'\to k}^{(1)}, \ldots,  V_{k'\to k}^{(\kappaR)}\colon k' \in \mathcal{I}_{\odd}, \;k\in\mathcal{I}_{\even}\Big\}.
\end{equation}

Our converse proof is based on the following observations:
\begin{itemize}
\item[i)] From $\mathbf V_{\even\to\odd}$ and $\mathbf Y_\odd^n$ it is possible to reconstruct 
\begin{equation}
\left\{\mathbf{V}_{\all \to k} \colon  k\in\mathcal{I}_{\odd}\right\},
\end{equation}
i.e., all  cooperation messages sent to odd-indexed receivers~$k\in\mathcal{I}_{\odd}$.

\item[ii)]From $\mathbf V_{\odd\to\even}$ and $\mathbf Y_\even^n$ it is possible to reconstruct 
\begin{equation}
\left\{\mathbf{V}_{\all\to k}\colon k\in\mathcal{I}_{\even}\right\},
\end{equation}
i.e., all  cooperation messages sent to even-indexed receivers~$k\in \mathcal{I}_{\even}$.


\item[iii)]
From $\textbf U_{\odd\to\even}$ and $\mathbf M_\even$ it is possible to reconstruct 
\begin{equation}
\left\{ \textbf{U}_{\all\to k} \colon k\in\mathcal{I}_{\even}\right\},
\end{equation}
i.e., all cooperation messages sent to even-indexed transmitters~$k\in\mathcal{I}_{\even}$.
\item[iv)] From $\textbf U_{\odd\to\even}$ and $\mathbf M_\even$ it is possible to reconstruct $\mathbf X_{\even}^n$.
\item[v)] From $\mathbf X_\even^n$,  $\mathbf Y_\even^n$, and $\mathbf G^n_{\noises,\even}$ it is possible to reconstruct $\mathbf Y_\odd^n$.
\end{itemize}

We now proceed to prove the converse:
\begin{IEEEeqnarray}{rCl}\lefteqn{
\sum_{k=1}^K R_{k} -\frac{\epsilon}{n}}\nonumber \\ &\leq &\frac{1}{n} I\Big(\mathbf M_{\even}; \mathbf{Y}_{\even}^n, \mathbf{V}_{\odd \to \even} | 
\mathbf G^n_{\noises,\even}\Big) \nonumber \\ 
& & + \frac{1}{n} I\Big(\mathbf M_{\odd}; \mathbf{Y}_{\odd}^n, \mathbf{V}_{\even \to \odd} | \mathbf M_{ \even}, \mathbf G^n_{\noises,\even}\Big)\nonumber \\
&\leq &\frac{1}{n} I\Big(\mathbf M_{\even}; \mathbf{Y}_{\even}^n, \mathbf{V}_{\odd\to \even} | 
\mathbf G^n_{\noises,\even}\Big) \nonumber \\ 
& & + \frac{1}{n} I\Big(\mathbf M_{\odd}, \mathbf U_{\odd \to \even}; \mathbf{Y}_{\odd}^n, \mathbf{Y}_{\even}^n | \mathbf M_{ \even}, \mathbf G^n_{\noises,\even}\Big)\nonumber \\
&= & {\frac{1}{n} I\Big(\mathbf M_{\even};\mathbf{V}_{\odd \to \even} |  \mathbf{Y}_{\even}^n, 
\mathbf G^n_{\noises,\even}\Big)}
 \nonumber \\ 
& & +{ \frac{1}{n} I\Big( \mathbf U_{\odd \to \even}  ; \mathbf{Y}_{\odd}^n, \mathbf{Y}_{\even}^n | \mathbf M_{ \even}, \mathbf G_{\noises,\even}^n\Big)}
\nonumber \\
& & + \frac{1}{n} I\Big(\mathbf M_{\even};\mathbf{Y}_{\even}^n| 
\mathbf G^n_{\noises,\even}\Big) \nonumber \\ 
& & + \frac{1}{n} I\Big(\mathbf M_{\odd}; \mathbf{Y}_{\odd}^n, \mathbf{Y}_{\even}^n | \mathbf{U}_{\odd \to \even}, \mathbf M_{ \even}, \mathbf G^n_{\noises,\even}\Big)\nonumber \\\label{eq:L} \\
& = &  {\frac{1}{n} I\Big(\mathbf M_{\even};\mathbf{V}_{\odd \to \even} |  \mathbf{Y}_{\even}^n, 
\mathbf G^n_{\noises,\even}\Big)}
 \nonumber \\ 
& & +{ \frac{1}{n} I\Big( \mathbf U_{\odd \to \even}  ; \mathbf{Y}_{\odd}^n, \mathbf{Y}_{\even}^n | \mathbf M_{ \even}, \mathbf G_{\noises,\even}^n\Big)}
\nonumber \\
& &+  \frac{1}{n} I\Big(\mathbf M_{\even};\mathbf{Y}_{\even}^n| 
\mathbf G^n_{\noises,\even}\Big) \nonumber \\ 
& & + \frac{1}{n} I\Big(\mathbf M_{\odd};  \mathbf{Y}_{\even}^n | \mathbf U_{\odd \to \even},\mathbf  M_{ \even}, \mathbf G^n_{\noises,\even}\Big) \nonumber \\
&= & {\frac{1}{n} I\Big(\mathbf M_{\even};\mathbf{V}_{\odd \to \even} |  \mathbf{Y}_{\even}^n, 
\mathbf G^n_{\noises,\even}\Big)}
 \nonumber \\ 
& & +{ \frac{1}{n} I\Big( \mathbf U_{\odd \to \even}  ; \mathbf{Y}_{\odd}^n, \mathbf{Y}_{\even}^n | \mathbf  M_{ \even}, \mathbf  G_{\noises,\even}^n\Big)}
\nonumber \\
& &  
 \underbrace{+\frac{1}{n}I\Big( \mathbf {Y}_{\even}^n ;   \mathbf U_{\odd \to \even},\mathbf  M_{ \even}, \mathbf M_{\odd}\big|\mathbf  G^n_{\noises,\even}\Big)}_{\leq \frac{K}{2} \cdot \frac{1}{2}\log(1+(1+\alpha_{\max})^2P)}
 \nonumber \\
& & \underbrace{- \frac{1}{n} 
I\Big(\mathbf{Y}_{\even}^n; \mathbf U_{\odd \to \even}\big|\mathbf  M_\even,
\mathbf G^n_{\noises,\even}\Big) }_{\leq 0}\nonumber \\ 
& \leq &  {\frac{1}{n} I\Big(\mathbf M_{\even};\mathbf {V}_{\odd \to \even} |  \mathbf {Y}_{\even}^n, 
\mathbf G^n_{\noises,\even}\Big)}
 \nonumber \\ 
& & +{ \frac{1}{n} I\Big( \mathbf  U_{\odd \to \even}  ; \mathbf {Y}_{\odd}^n, \mathbf {Y}_{\even}^n | \mathbf M_{ \even},\mathbf  G_{\noises,\even}^n\Big)}
\nonumber \\
 & & + \frac{K}{2} \cdot\frac{1}{2}\log(1+(1+\alpha_{\max})^2P). \label{eq:upperbound1}
\end{IEEEeqnarray}
Above steps are justified as follows: 
\begin{itemize}
\item The first inequality holds by Fano's inequality, by the independence of $\mathbf  G^n_{\noises,\even}$, $\mathbf  M_\even$ and $\mathbf  M_{\odd}$, and by our previous observations~i) and ii).
\item The second inequality holds because
$\mathbf V_{\even\to\odd}$ can be computed from $(\mathbf Y_\even^n,\mathbf  Y_\odd^n)$ and $\mathbf U_{\odd\to\even}$ can be computed from $(\mathbf M_\even,\mathbf  M_\odd)$.
\item The first equality holds by the chain rule for mutual information. 
\item The second equality holds by our previous observations~iv) and v).
\item The third equality holds by the definition of mutual information and by rearranging terms.
\item The last inequality holds by the nonnegativity of mutual information and by Inequality
\begin{IEEEeqnarray}{rCl}
\lefteqn{
I\Big( \mathbf {Y}_{\even}^n ;   \mathbf U_{\odd \to \even},\mathbf  M_{ \even}, \mathbf M_{\odd}\big|\mathbf  G^n_{\noises,\even}\Big)} \qquad \nonumber \\
& \leq&\frac{K}{2} \cdot\frac{1}{2}\log\left(1+(1+\alpha_{\max})^2P\right). 
\end{IEEEeqnarray} 
 \end{itemize}

 Observations~i)--iv)  hold also when the subscripts ``$\even$" and ``$\odd$" are exchanged. The same is true for the above sequence of inequalities leading to \eqref{eq:L}. Thus, 
\begin{IEEEeqnarray}{rCl}
\lefteqn{\sum_{k=1}^{K} R_{k}-\frac{\epsilon}{n}}  \nonumber \\
& \leq &  {\frac{1}{n} I\Big(\mathbf M_{\odd};\mathbf{V}_{\even \to \odd} |  \mathbf{Y}_{\odd}^n, 
\mathbf G^n_{\noises,\odd}\Big)}
 \nonumber \\ 
& & +{ \frac{1}{n} I\Big( \mathbf U_{\even \to \odd}  ; \mathbf{Y}_{\even}^n, \mathbf{Y}_{\odd}^n | \mathbf M_{ \odd}, \mathbf G_{\noises,\odd}^n\Big)}
\nonumber \\
& &+  \frac{1}{n} I\Big(\mathbf M_{\odd};\mathbf{Y}_{\odd}^n| 
\mathbf G^n_{\noises,\odd}\Big) \nonumber \\ 
& & + \frac{1}{n} I\Big(\mathbf M_{\even}; \mathbf {Y}_{\even}^n,  \mathbf{Y}_{\odd}^n | \mathbf U_{\even \to \odd},\mathbf  M_{ \odd}, \mathbf G^n_{\noises,\odd}\Big).\IEEEeqnarraynumspace
\end{IEEEeqnarray}
Now, since from  $\mathbf X_\odd^n$,  $\mathbf Y_\odd^n$, and $\mathbf G^n_{\noises,\odd}$ it is possible to reconstruct $Y_2^n, \ldots, Y_{K-2}^n$ (but not $Y_K^n$), and by using the definition of mutual information and rearranging terms:
\begin{IEEEeqnarray}{rCl}
\lefteqn{\sum_{k=1}^{K} R_{k}-\frac{\epsilon}{n}} \quad \nonumber \\
& \leq &  {\frac{1}{n} I\Big(\mathbf M_{\odd};\mathbf{V}_{\even \to \odd} |  \mathbf{Y}_{\odd}^n, 
\mathbf G^n_{\noises,\odd}\Big)}
 \nonumber \\ 
& & +{ \frac{1}{n} I\Big( \mathbf U_{\even \to \odd}  ; \mathbf{Y}_{\even}^n, \mathbf{Y}_{\odd}^n | \mathbf M_{ \odd}, \mathbf G_{\noises,\odd}^n\Big)}
\nonumber \\
& &+  \frac{1}{n} I\Big(\mathbf M_{\odd};\mathbf{Y}_{\odd}^n| 
\mathbf G^n_{\noises,\odd}\Big) \nonumber \\ 
& & + \frac{1}{n} I\Big(\mathbf M_{\even};  {Y}_{K}^n, \mathbf{Y}_{\odd}^n | \mathbf U_{\even \to \odd},\mathbf  M_{ \odd}, \mathbf G^n_{\noises,\odd}\Big)\nonumber\\
& = &  {\frac{1}{n} I\Big(\mathbf M_{\odd};\mathbf{V}_{\even \to \odd} |  \mathbf{Y}_{\odd}^n, 
\mathbf G^n_{\noises,\odd}\Big)}
 \nonumber \\ 
& & +{ \frac{1}{n} I\Big( \mathbf  U_{\even \to \odd}  ; \mathbf {Y}_{\odd}^n, \mathbf {Y}_{\even}^n | \mathbf M_{ \odd},\mathbf  G^n_{\noises,\odd}\Big)}
\nonumber \\ 
& & \underbrace{+\frac{1}{n} I\Big(\mathbf M_{\odd}, \mathbf  U_{\even \to \odd}, \mathbf M_{\even} ;\mathbf{Y}_{\odd}^n| 
\mathbf G^n_{\noises,\odd}\Big)}_{\leq \frac{K}{2} \cdot \frac{1}{2}\log(1+(1+\alpha_{\max})^2P)} \nonumber \\ 
& & \underbrace{- \frac{1}{n} I\Big(\mathbf U_{\even \to \odd}; \mathbf{Y}_{\odd}^n | \mathbf  M_{ \odd}, \mathbf G^n_{\noises,\odd}\Big)}_{\leq 0}\nonumber\\
& & \underbrace{+ \frac{1}{n} I\Big(\mathbf M_{\even};  {Y}_{K}^n| \mathbf{Y}_{\odd}^n , \mathbf U_{\even \to \odd},\mathbf  M_{ \odd}, \mathbf G^n_{\noises,\odd}\Big)}_{\leq \frac{1}{2}\log(1+(1+\alpha_{\max})^2P)}\nonumber \\
& \leq &  {\frac{1}{n} I\Big(\mathbf M_{\odd};\mathbf{V}_{\even \to \odd} |  \mathbf{Y}_{\odd}^n, 
\mathbf G^n_{\noises,\odd}\Big)}
 \nonumber \\ 
& & +{ \frac{1}{n} I\Big( \mathbf  U_{\even \to \odd}  ; \mathbf {Y}_{\odd}^n, \mathbf {Y}_{\even}^n | \mathbf M_{ \odd},\mathbf  G^n_{\noises,\odd}\Big)}
\nonumber \\ 
& & +\left(\frac{K}{2}+1\right) \frac{1}{2}\log\left(1+(1+\alpha_{\max})^2P\right).\IEEEeqnarraynumspace\label{eq:upperbound2}
\end{IEEEeqnarray}

In the following we combine bounds~\eqref{eq:upperbound1} and \eqref{eq:upperbound2}. To this end, notice that the transmitter-side conferencing constraint implies 
\begin{IEEEeqnarray}{rCl}
\lefteqn{\frac{1}{n} I\Big( \mathbf  U_{\odd \to \even}  ; \mathbf {Y}_{\odd}^n, \mathbf {Y}_{\even}^n | \mathbf M_{ \even},\mathbf  G_{\noises}^n\Big) }\quad \nonumber \\ 
& +& 
 \frac{1}{n} I\Big( \mathbf  U_{\even \to \odd}  ; \mathbf {Y}_{\odd}^n, \mathbf {Y}_{\even}^n | \mathbf M_{ \odd},\mathbf  G_{\noises}^n\Big) \nonumber \\
 && \; \leq K\mu_\Tx(\tL+\tR)  \frac{1}{2}\log(1+P).\IEEEeqnarraynumspace\label{eq:prelogtx}
\end{IEEEeqnarray}
and the receiver-side conferencing constraint implies
\begin{IEEEeqnarray}{rCl}
\lefteqn{\frac{1}{n} I\Big(\mathbf M_{\even};\mathbf {V}_{\odd \to \even} |  \mathbf {Y}_{\even}^n, 
\mathbf G_{\noises}^n\Big)} \quad \nonumber \\ 
& +& 
  \frac{1}{n} I\Big(\mathbf M_{\odd};\mathbf {V}_{\even \to \odd} |  \mathbf {Y}_{\odd}^n, 
\mathbf G_{\noises}^n\Big) \nonumber \\
 && \; \leq 2K\mu_\Rx  \frac{1}{2}\log(1+P).\IEEEeqnarraynumspace\label{eq:prelogrx}
\end{IEEEeqnarray}
Adding up bounds \eqref{eq:upperbound1} and \eqref{eq:upperbound2} and dividing the result by 2, in view of \eqref{eq:prelogtx} and \eqref{eq:prelogrx}, we obtain for any even value of $K$:
\begin{IEEEeqnarray}{rCl}\label{eq:evenK}
\lefteqn{ \sum_{k=1}^K R_k\leq \frac{1}{2}\big(K+1+ 2K\mu_\Tx+2K\mu_{\Rx}\big)} \hspace{2cm} \nonumber \\ &&\cdot  \frac{1}{2}\log\big(1+(1+\alpha_{\max})^2P\big).\hspace{2cm}
\end{IEEEeqnarray}
The same bound can also be obtained for odd values of $K$. After dividing by $\frac{K}{2} \log (1+P)$ and letting first $P\to \infty$ and then $K\to \infty$,  bound~\eqref{eq:evenK}  establishes the desired upper bound on the per-user MG in \eqref{eq:converse_infinity}.

\section{Summary and Conclusion}
We quantify how the asymptotic MG per-user of Wyner's soft-handoff interference network increases with the number of permitted Tx- and Rx-conferencing rounds. We identify two regimes. When the conferencing links are of low rate, then the asymptotic MG per-user does not depend on the number of permitted conferencing rounds; a single round of non-interactive conferencing suffices. When the conferencing links are of high rate, then every additional conferencing round increases the asymptotic per-user MG. For certain system parameters there is a third intermediate regime, where a single conferencing round is suboptimal, but a finite number of rounds suffices  to achieve the asymptotic MG per-user.

Determining the smallest number of conferencing rounds that attains the asymptotic MG per user is of practical interest, because it allows to limit implementation complexity. 

Intuitively, increasing the number of Tx- and Rx-conferencing rounds in an interference network can be beneficial, because 
information about a given transmit message or about a given receive signal can be spread over a larger part of the network. This is important for iterative interference mitigation techniques  (like successive dirty-paper coding at the transmitters and successive interference cancellation at the receivers), where interference-mitigation information 
 precisely needs to propagate over the network.

To avoid propagating interference beyond what can be mitigated with the number of permitted conferencing rounds, in our scheme we periodically deactivate transmitters. This splits the large network into smaller subnets. 
Over each of these subnets we employ a coding scheme that smartly combines transmitter and receiver interference-mitigation techniques so as to allow to keep the subnets as large as possible, and thus minimize the number of  deactivated transmitters. 

The conferencing protocols and interference-mitigation techniques that we use in the subnets, are inspired by Ntranos, Maddah-Ali, and Caire~\cite{ntranosmaddah-alicaire14-1}. Transmitters describe quantised versions of transmit signals over the Tx-conferencing links to their left- or right-neighbours, and these neighbours apply dirty-paper coding to mitigate the interference signals described over the conferencing links. Receivers send decoded messages over Rx-conferencing links to their left- or right-neighbours. These neighbours then reconstruct the transmit signals corresponding to the conferenced messages, and subtract these interferences from their receive signals. 

In general, the described conferencing  strategies are strictly better than conferencing messages at the transmitter-side and quantised versions of receive signals at the receiver-side. The advantage of this latter conferencing strategy however is that it can be applied also in oblivious setups (like for example in C-RANs) where the codebooks are not known during the Tx- and Rx-conferencing phases. When only a single conferencing round is permitted at the transmitter and the receiver side, than the two conferencing strategies are equivalent. In this sense, our results also provide an estimate about the loss in asymptotic MG per user in  oblivious setups.

%

\appendices

\section{Detailed description and analysis of the coding schemes in a subnet} \label{app3}

We describe and analyse the random coding argument that we employ for the first subnet. We will construct random codebooks, which are revealed to all transmitters and receivers before communication starts. The probabilities of decoding errors that we present in our analysis are \emph{average} probabilities of error, where the average is taken over the random source messages, the random channel realisations and the random choices of the codebooks. We shall identify conditions under which these average probabilities of decoding errors tend to 0  as the blocklength $n\to \infty$.  Standard arguments then imply that there must exist a deterministic choice of all the codebooks such that the average probabilities of decoding errors (now averaged only over the source messages and the channel realisations) tend to 0 as $n\to \infty$.

Recall that in the first subnet we transmit messages $M_{\i}, \ldots, M_{\i+\beta-\kappaR-2}, M_{\i+\beta-\kappaR-1}^{(\textnormal{Tx})}, M_{\i+\beta-\kappaR-1}^{(\textnormal{Rx})}, M_{\i+\beta-\kappaR}, \ldots, M_{\i+\beta-2}$, and that these messages are partitioned into four groups, see subsection~\ref{sec:subnetcom} and figure~\ref{Fig:Subnet1}. We explain transmission of these four groups separately. For ease of exposition, we start with group~1, followed by group~2, then group~4, and finally group~3. 
  
\subsection{Transmission of source messages $M_{\i}, \ldots, M_{\i+\kappaR}$ (messages in group~1)}\label{sec:Group1}

For each $k \in \{\i,\i+1,\ldots, \i+\kappaR\}$, let 
\begin{equation*}
\s{C}_{\text{P2P},k} := \Big\{X^n_{k}(m) = \big(X_{k,1}(m)\ \ldots\ X_{k,n}(m)\big)\Big\}_{m = 1}^{\lfloor 2^{nR_k}\rfloor}
\end{equation*}
be a random Gaussian codebook of rate $R_k$ with codewords of length $n$ drawn iid as $X_k \sim \s{N}(0,P)$.\footnote{To be precise, in order to satisfy the power constraint $P$ in \eqref{eq:power}, the variance of $X_k$ needs to be chosen slightly smaller than $P$. This is a technicality that we will ignore for ease of exposition.}  Given  source message~$M_k$, Tx~$k$ sends the corresponding codeword $X^n_{k}(M_k)$ over the channel. 


Consider Rx~$\i $ (the first receiver in group one). Tx~$\i-1$ has been deactivated, so the channel output at Rx~$\i$ is
\begin{equation}\label{eq:kappaoutput}
Y_{\i}^n = X_{\i}^n(M_\i) + Z_{\i}^n,
\end{equation}
where the addition is understood to be symbol by symbol.  Rx~$\i$ looks through the codebook $\s{C}_{\text{P2P},\i}$ for a unique index ${m}^*$ such that $X_{\i}^n({m}^*)$ and $Y_{\i}^n$ are jointly typical \cite{elgamalkim10}. If successful, Rx~$\i$ declares $\h{M}_\i ={m}^*$, otherwise it declares $\h{M}_\i = 1$.  By standard arguments \cite{elgamalkim10}, if
\begin{equation}\label{eq:RRR1}
R_ \i < \frac{1}{2} \log (1+P),
\end{equation}
then $\Pr[\h{M}_\i \neq M_\i] \to 0$ as $n\to \infty$.

Rx~$\i$ sends its estimate $\h{M}_\i$ of $M_\i$ to Rx~$\i+1$ during the first Rx-conferencing round:
\begin{equation}\label{eq:conferencing1}
V\bs{1}_{\i \to \i + 1} := \h{M}_\i.
\end{equation}
(This will be the only conferencing message that Rx~$\i$ sends). 

Rx~$\i + 1$ estimates the \emph{interference} from Tx~$\i$ to be $\alpha_{\i + 1}\ X^n_\i(\h{M}_\i)$ and computes
\begin{equation}\label{Eqn:Rx-Interference-Cancellation}
\h{Y}^n_{\i + 1} := Y^n_{\i + 1} - \alpha_{\i + 1} X^n_\i(\h{M}_\i).
\end{equation}
If Rx~$\i$ decoded correctly, $\h{M}_\i = M_\i$, then $X^n_\i(\hat{M}_\i)= X^n_\i({M}_\i)$ and
\begin{equation*}
\h{Y}^n_{\i + 1}= X^n_{\i+1}(M_{\i + 1}) + Z^n_{\i+1}.
\end{equation*} 
Rx~$\i + 1$ looks through its codebook $\s{C}_{\text{P2P},\i+1}$ for a unique index ${m}^*$ such that $X_{\i+1}^n({m}^*)$ and $\h{Y}_{\i+1}^n$ are jointly typical. If successful, it declares $\h{M}_{\i+1} ={m}^*$, otherwise it declares $\h{M}_{\i+1} = 1$. If
\begin{equation}\label{eq:r2}
R_{\i + 1} < \frac{1}{2} \log(1 + P),
\end{equation}
then $\Pr[\h{M}_{\i+1} \neq M_{\i+1} | \h{M}_\i = M_\i] \to 0$ as $n\to \infty$. 

Rx~$\i + 1$ sends its estimate $\h{M}_{\i+1}$ to Rx~$\i+2$ during the second Rx-conferencing round:
\begin{equation}\label{eq:conferencing2}
V\bs{2}_{\i + 1 \to \i + 2} = \h{M}_{\i + 1}.
\end{equation}  

The same process is repeated for  receivers~$\i+2, \ldots, \i+\kappaR$ in increasing order, see also Figure~\ref{Fig:GroupG1}. The only difference concerns the Rx-conferencing messages:  The right-most receiver Rx~$\i+\kappaR-1$ does not send any conferencing message at all.  Every other receiver~$\i+j$, for $j\in\{2,\ldots, \kappaR-1\}$,  has to wait until Rx-conferencing round $j+1$ to send its conferencing message $\h{M}_{\i+j}$:
\begin{equation}
V\bs{j+1}_{\i + j \to \i + j+1} = \h{M}_{\i + j}, \qquad j\in\{2,\ldots, \kappaR-1\}.
\end{equation}

The next lemma follows by iteratively applying the arguments we used to derive~\eqref{eq:RRR1} and \eqref{eq:r2} and accounting for \eqref{eq:conferencing1} and \eqref{eq:conferencing2}.

\bigskip
\begin{lemma}\label{Lem:Group1}
Source messages $M_{\i}, \ldots, M_{\i+\kappaR}$ are successfully decoded  with probability tending to 1 as $n\to \infty$, whenever
 \begin{equation*}
 R_k < \frac{1}{2} \log\left(1+ P\right),
 \qquad
\forall \ k \in \{\i, \ldots, \i+\kappaR\},
 \end{equation*}  
 and 
 \begin{equation*}
R_\Tx > R_{k}, \qquad \forall k\in \{\i, \ldots, \i+\kappaR-1\}.
\end{equation*} 

 \end{lemma}
 

 \subsection{Transmission of source messages $M_{\i+\kappaR+1}, \ldots, M_{\i+\kappaR+\kappaT}$ (messages in group~2)}\label{sec:Group2}
 
Let us temporarily fix $k = \i + \kappaR + 1$ to simplify notation. Consider Tx~$k - 1$ (the last transmitter in group one), and recall that its channel codeword $X^n_{k-1}(M_{k-1})$ was chosen from an iid Gaussian codebook of power $P$.  To help facilitate communication at Tx~$k$ (the first transmitter of group 2), Tx~$k-1$ sends a rate-$1/2\log(1+P)$ quantisation of its transmitted signal $X^n_{k-1}(M_{k-1})$ to Tx~$k$. 

Let 
\begin{equation*}
\h{X}_{k-1} \sim \s{N}\left(0,\frac{P^2}{1+P}\right)
\quad
\text{and}
\quad 
Z^\dag_{k-1} \sim \s{N}\left(0,\frac{P}{1+P}\right).
\end{equation*}
so that 
\begin{equation*}
X_{k-1} := \h{X}_{k-1} + Z^\dag_{k-1} \sim \s{N}(0,P).
\end{equation*}
Construct a random quantisation codebook 
\begin{equation*}
\s{C}_{\text{RD}, k-1} 
:= \Big\{ 
\h{X}_{k-1}^n(u) 
= \big(\h{X}_{k-1,1}(u)\ \ldots\ \h{X}_{k-1,n}(u)\big)   \Big\}_{u = 1}^{\lfloor 2^{n R_\Tx}\rfloor }
\end{equation*}
 with codewords of length $n$ drawn iid as $\hat{X}_{k-1}$.
 
  Tx~$k-1$ takes its channel codeword $X^n_{k-1}(M_{k-1})$ and looks through $\s{C}_{\text{RD}, k-1}$ for a unique index ${u}^*$ such that $X^n_{k-1}(M_{k-1})$ and $\h{X}^n_{k-1}({u}^*)$ are  jointly typical. If successful, Tx~$(k-1)$ sends the index 
 \begin{equation}
 U\bs{1}_{k-1 \to k} ={u}^*
 \end{equation}
  to Tx~$k$ during Tx-conferencing round 1; otherwise it sends $U\bs{1}_{k-1 \to k} = 1$. (This is the only conferencing message Tx~$k-1$ sends.)

  Let $\mathcal{Q}_{k-1}$ denote the event that the described quantisation is successful, i.e., that there was a unique index ${u}^*$. Standard arguments show that whenever the Tx-conferencing rate satisfies
\begin{equation}\label{eq:conferencing3}
R_\Tx> \frac{1}{2} \log (1 + P), 
\end{equation}
then  $\Pr[\mathcal{Q}_{k-1}] \to 1$ as $n\to\infty$.

Now consider Tx/Rx pair~$k$.
%
%
Rx~$k$ observes the channel outputs
\begin{equation*}
Y_k^n = \alpha_k X^n_{k-1}(M_{k-1}) + X^n_k + Z^n_k,
\end{equation*}
which can be rewritten as
\begin{equation}\label{eq:outputY_k}
Y_k^n = 
\underbrace{X_k^n}_{\text{channel input}} 
+ \underbrace{\alpha_k  \h{X}^n_{k-1}(U\bs{1}_{k-1 \to k})}_{\text{channel state $S_k^n$ known at Tx $k$ } }
+ \quad \underbrace{
\alpha_{k}\big(X_{k-1}^n(M_k) -  \h{X}^n_{k-1}(U\bs{1}_{k-1 \to k})\big) + Z_k^n }_{\text{additive noise }\tilde{Z}_k^n}.
\end{equation}
%
After obtaining the  conferencing message $U\bs{1}_{k-1 \to k}$, Tx~$k$ reconstructs $\hat{X}_{k-1}( U\bs{1}_{k-1 \to k})$. It then encodes its source message $M_k$ using \emph{dirty-paper coding over spheres} \cite{cohenlapidoth02-2},\cite[Section~V]{brosslapidothwigger12}\footnote{Standard dirty-paper coding and its analysis are not sufficient because 
 the noise sequence $\tilde{Z}_k^n$ is neither iid (not even when averaged over all codebooks) nor independent of the state sequence $S_k^n$.
 
  In fact, since the state-sequence $S_k^n$ is not  uniform over a sphere, the dirty-paper  encoding over spheres needs to be extended as described in the proof of Remark~III-5 in \cite{brosslapidothwigger12}. The analysis of this extended {dirty-paper coding over spheres} only requires that the normalised lengths of the noise and the state sequences are approximately constant and the noise and state sequences are approximately orthogonal \cite{brosslapidothwigger12}. Given event $\mathcal{Q}_{k-1}$,  our setup satisfies these conditions:   $\frac{1}{n}\big\|\tilde{Z}_{k}^n\big\|^2 \ 
  \longrightarrow\ 1+\alpha_k^2\frac{P}{P+1}$,  \; $\frac{1}{n}\big\|{S}_{k}^n\big\|^2\longrightarrow \alpha_k^2 \frac{P^2}{P+1}$, \; and  $\frac{1}{n} <\tilde{Z}_{k}^n, S_k^n> \ \longrightarrow\ 0$, where convergence is in probability everywhere.} 
of power $P$ and designed for the channel in \eqref{eq:outputY_k} with additive state sequence 
 \begin{equation}
 S_k^n:=\alpha_k\h{X}^n_{k-1}(U\bs{1}_{k-1 \to k})
 \end{equation} 
 and additive noise sequence
\begin{equation*}
\tilde{Z}_k^n:=\alpha_{k}\big(X_{k-1}^n(M_k) -  \h{X}^n_{k-1}(U\bs{1}_{k-1 \to k})\big) + Z_k^n.
\end{equation*}  

Rx~$k$ uses dirty-paper decoding over spheres as described in \cite[Section~V-C]{brosslapidothwigger12}.
Adapting Corollary~IV.2  in 
\cite{brosslapidothwigger12} similarly as in Remark~IV.4 of \cite{brosslapidothwigger12}, and in view of footnote 10, we obtain the following: If Tx~$k$ sends its source message $M_k$ to Rx~$k$ using  dirty-paper coding over spheres~\cite[Section~V]{brosslapidothwigger12} of power $P$ and designed for a channel with normalised noise variance $N=1+\alpha_k^2\frac{P}{P+1}$ and interference sequence $S_k^n$, and if the rate of the source message
\begin{equation}\label{eq:prelog16}
R_k < \frac{1}{2} \log\left( 1+ \frac{P}{1+\alpha_k^2\frac{P}{P+1}}\right), %
\end{equation}
then $\Pr[\h{M}_{k} \neq M_{k}|\s{Q}_{k-1}] \to 0$ as $n\to \infty$.\\

Let now $k=\i+\kappaR+2$ and consider Tx~$k-1$. (This is the first transmitter in group 2 that we also considered in the previous dirty-paper coding step.) It facilitates communication at Tx~$k$ (the second transmitter in group 2) by sending it a rate-$\frac{1}{2}\log(1+P)$ quantisation of its input signal $X_{k-1}^n$. Since this input signal was
 produced by the dirty-paper coding over spheres in \cite[Section~V]{brosslapidothwigger12}), we use a quantisation codebook 
\begin{equation*}
\s{C}_{\text{RD}, k-1} 
:= \Big\{ 
\h{X}_{k-1}^n(u) 
   \Big\}_{u = 1}^{\lfloor 2^{n R_\Tx}\rfloor}
\end{equation*} with codewords $\h{X}_{k-1}^n(u)$ that are picked iid uniformly over the surface of an $n$-dimensional sphere of radius $\sqrt{n \textnormal{Var}\big(\hat{X}_{k-1}\big)}=\sqrt{n\frac{P^2}{1+P}}$. 
 Quantisation is as follows. Tx~$k-1$  looks through $\s{C}_{\text{RD}, k-1}$ for the vector $\h{X}_{k-1}^n(u^*)$ whos angle with $X^n_{k-1}$ is closest to~$\sqrt{\frac{P}{P+1}}$:
 \begin{equation}
 u^* =\text{argmin}_{u\in\{1,\ldots, 2^{nR_\Tx}\}} \bigg| \angle\big(\h{X}_{k-1}^n(u),{X}_{k-1}^n\big)- \sqrt{\frac{P}{P+1}}\bigg|.
 \end{equation} 
Tx~$k-1$ sends the index $u^*$ over the conferencing link to Tx~$k$. It does so during the second Tx-conferencing round: 
%
\begin{equation*}
U\bs{2}_{k-1\to k} =u^*.
\end{equation*}
  By standard arguments (see e.g., proof of (134) in \cite{lapidothtinguelyMAC}), whenever 
 \begin{equation}\label{eq:rr11}
  R_\Tx> \frac{1}{2}\log(1+P),
 \end{equation}
then
 \begin{equation}\label{eq:co}
 \bigg| \angle\big(\h{X}_{k-1}^n(u^*),{X}_{k-1}^n\big)- \sqrt{\frac{P}{P+1}}\bigg| \longrightarrow 0 \qquad \textnormal{in probability}.
 \end{equation}



Tx~$k$ reconstructs the quantised signal $\hat{X}_{k-1}^n(U\bs{2}_{k-1\to k})$,  and encodes its source message $M_k$ using dirty-paper coding over spheres of power $P$ and for a channel with normalised noise variance $N=1+\alpha_k^2\frac{P}{P+1}$ and interference sequence $S_k^n:= \alpha_{k} \hat{X}_{k-1}^n(U\bs{2}_{k-1\to k})$. 

Receiver~$k$ applies  dirty-paper coding over spheres to decode message $M_k$ from its output sequence
\begin{equation*}
Y_k^n = 
\underbrace{X_k^n}_{\text{channel input}} 
+ \underbrace{\alpha_k  \h{X}^n_{k-1}(U\bs{1}_{k-1 \to k})}_{\text{channel state $S_k^n$ known at Tx $k$ } }
+ \quad \underbrace{
\alpha_{k}\big(X_{k-1}^n(M_k) -  \h{X}^n_{k-1}(U\bs{1}_{k-1 \to k})\big) + Z_k^n }_{\text{additive noise }\tilde{Z}_k^n}.
\end{equation*}

Since given \eqref{eq:rr11} the noise sequence $\tilde{Z}_k^n$ and state sequence $S_k^n$ satisfy again the convergence conditions in footnote~10, extending 
 Corollary~IV.2 similarly to Remark~IV.4 (both in \cite{brosslapidothwigger12}), we obtain that if  \eqref{eq:rr11} and
 \begin{equation}\label{eq:prelog17}
R_k < \frac{1}{2} \log\left( 1+ \frac{P}{1+\alpha_k^2\frac{P}{P+1}}\right) %
\end{equation}
hold, then $\Pr[\h{M}_{k} \neq M_{k}] \to 0$ as $n\to \infty$.\\
 
 The same procedure can be repeated for $k=\i+\kappaR+3,\ldots, \i+\kappaR+\kappaT$, see also Figure~\ref{Fig:LeftDirtyPaperCoding}. 
 
 The next lemma follows by iteratively applying the arguments we used to derive~\eqref{eq:prelog16} and \eqref{eq:prelog17}.

 \bigskip
\begin{lemma}\label{Lem:Group2}
Source messages $M_{\i+\kappaR+1}\ldots, M_{\i+\kappaR+\kappaT}$ are successfully decoded  with probability tending to 1 as $n\to \infty$, whenever
 \begin{equation*}
 R_k < \frac{1}{2} \log\left(1+  P\right), \qquad \forall k \in \{\i+\kappaR+1,\ldots, \i+\kappaR+\kappaT\}
  \end{equation*}  
  and 
   \begin{equation*}
  R_\Tx> \frac{1}{2}\log(1+P).
 \end{equation*}
\end{lemma}
 
\subsection{Transmission of source messages $M_{\i+\beta-\kappaR-1}^{\textnormal{Rx}}, M_{\i+\beta-\kappaR}, \ldots, M_{\i+\beta-2}$ (messages in group~4)}\label{sec:Group4}

For each $k \in \{\i+\beta-\kappaR-1,\ldots, { \i+\beta -2}\}$, let 
\begin{equation*}
\s{C}_{\text{P2P},k} := \Big\{\Xi^n_{k}(m) = \big(\Xi_{k,1}(m)\ \ldots\ \Xi_{k,n}(m)\big)\Big\}_{m = 1}^{\lfloor 2^{nR_k} \rfloor}.
\end{equation*}
be a random Gaussian codebook of rate $R_k$ with codewords of length $n$ drawn iid as $\Xi_k \sim \s{N}(0,\alpha_{k+1}^2 P)$. Given  source message $M_k$,\footnote{Whenever we write $M_k$ or $R_k$ for $k=\i+\beta-\kappaR-1$ in this subsection we actually mean $M_{\i+\beta-\kappaR-1}^{\textnormal{Rx}}$ and $R_{\i+\beta-\kappaR-1}^{\textnormal{Rx}}$. We do not write the latter for ease of exposition.} Tx~$k$ picks the corresponding codeword $\Xi^n_{k}(M_k)$ and transmits the scaled sequence
\begin{equation}\label{eq:inputscaling}
X_{k}^n = \alpha_{k+1}^{-1} \Xi_{k}^n(M_k).
\end{equation}

Transmission of each source message $M_k$ goes over the ``diagonal path"
\begin{equation*}
\textnormal{Tx}~k \quad \longrightarrow\quad  \textnormal{Rx}~k+1\quad  \longrightarrow\quad  \textnormal{Rx}~k.
\end{equation*}
In fact, Rx~$k+1$ decodes source message $M_k$ and describes its guess over the conferencing link to Rx~$k-1$, which then declares this message.

Consider Rx~$\i+\beta-1$ (the last receiver in the subnet). Since in our scheme Tx~$\i+\beta-1$ is deactivated, Rx~$\i+\beta-1$ observes 
\begin{IEEEeqnarray}{rCl}\label{eq:channel_last}
 Y_{\i+\beta-1}^n&=&\alpha_{\i+\beta-1} X_{\i+\beta-2}^n + Z_{\i+\beta-1}^n \nonumber \\
 &=&\Xi_{\i+\beta-2}^n(M_{\i+\beta-2 }) + Z_{\i+\beta-1}^n,
\end{IEEEeqnarray} 
where the second equality follows by \eqref{eq:inputscaling}. 

Rx $\i+\beta-1$ decodes source message $M_{\i+\beta-2 }$. It  looks through codebook $\s{C}_{\text{P2P},\i+\beta-2}$ for a unique index ${m}^*$ such that $\Xi_{\i+\beta-2}^n({m}^*)$ and ${Y}_{\i+\beta-1}^n$ are jointly typical. If successful it sets $\hat{M}_{\i+\beta-2}={m}^*$, otherwise it sets $\hat{M}_{\i+\beta-2}=1$. By standard arguments \cite{coverthomas06}, if 
\begin{equation}\label{eq:RRR4}
R_{\i+\beta-2} < \frac{1}{2}\log(1 + \alpha_{\i+\beta-1}^2P),
\end{equation}
then $\Pr[\h{M}_{\i+\beta-2} \neq M_{\i+\beta-2}] \to 0$ as $n\to \infty$. 

Rx~$\i+\beta-1$ sends the conferencing message
\begin{equation}\label{eq:Rxconfff}
V\bs{1}_{(\i+\beta-1) \to (\i+\beta-2)} = \h{M}_{\i+\beta-2}
\end{equation} to Rx~$\i+\beta-2$. (This is the only conferencing message Rx~$\i+\beta-1$ sends.) 

We next consider Rx~$\i+\beta-2$. (This is the receiver immediately left to the previously considered receiver, which obtained the conferencing message \eqref{eq:Rxconfff}.) It observes channel outputs 
\begin{IEEEeqnarray}{rCl}
Y_{\i+\beta-2}^n&=&X_{\i+\beta-2}^n+\alpha_{\i+\beta-2} X_{\i+\beta-3}^n+ Z_{\i+\beta-2}^n \nonumber \\
& = &\underbrace{\alpha_{{\i+\beta-1}}^{-1} \Xi_{\i+\beta-2}^n(M_{\i+\beta-2})}_{\textnormal{interference ``known" at receiver}}+\underbrace{\Xi_{\i+\beta-3}^n(M_{\i+\beta-3})}_{\textnormal{desired signal}}+ \underbrace{Z_{\i+\beta-2}^n}_{\textnormal{noise}}
\end{IEEEeqnarray} 
where the second equality follows again by \eqref{eq:inputscaling}. 

Rx~$\i+\beta-2$  has a guess of $M_{\i+\beta-2}$, see \eqref{eq:Rxconfff}, and thus an estimate about $\alpha_{{\i+\beta-1}}^{-1} \Xi_{\i+\beta-2}^n(M_{\i+\beta-2})$. It will cancel this ``interference" before decoding source message  $M_{\i+\beta-3}$.
Specifically, Rx~$\i+\beta-2$ first forms 
\begin{IEEEeqnarray}{rCl}
\hat{Y}_{\i+\beta-2}^n&=&{Y}_{\i+\beta-2}^n - \alpha_{{\i+\beta-1}}^{-1} \Xi_{\i+\beta-2}^n(\hat{M}_{\i+\beta-2}),
\end{IEEEeqnarray} 
and then looks through codebook $\s{C}_{\text{P2P},\i+\beta-2}$ for a unique index ${m}^*$ such that $\Xi_{\i+\beta-3}^n({m}^*)$ and $\h{Y}_{\i+\beta-2}^n$ are jointly typical. If successful it sets $\hat{M}_{\i+\beta-3}={m}^*$, otherwise it sets $\hat{M}_{\i+\beta-3}=1$. 

Notice that when  $\h{M}_{\i+\beta-2}={M}_{\i+\beta-2}$, then  
\begin{IEEEeqnarray}{rCl}
\hat{Y}_{\i+\beta-2}^n&=&\Xi_{\i+\beta-3}^n(M_{\i+\beta-3})+ Z_{\i+\beta-2}^n,
\end{IEEEeqnarray} 
and Rx~$\i+\beta-2$ can declare source message $M_{\i+\beta-3}$ based on an interference-free signal.

 By standard arguments \cite{coverthomas06}, $\Pr[\h{M}_{\i+\beta-3} \neq M_{\i+\beta-3}|\h{M}_{\i+\beta-2} = M_{\i+\beta-2}] \to 0$ as $n\to \infty$, whenever 
\begin{equation}\label{eq:r4}
R_{\i+\beta-3} < \frac{1}{2} \log(1 + \alpha_{\i+\beta-2}^2P).
\end{equation}

Rx~$\i+\beta-2$ sends the Rx-conferencing message
\begin{equation*}
V\bs{2}_{(\i+\beta-2) \to (\i+\beta-3)} = \h{M}_{\i+\beta-3}
\end{equation*} to its left neighbour. (This is its only conferencing message.) 

Finally, Rx~$\i+\beta-2$ declares the guess $\hat{M}_{\i+\beta-2}$ that it had obtained from its right neighbour \eqref{eq:Rxconfff}.\\

The same process is repeated for  receivers $\i+\beta-3, \i+\beta-4, \ldots, i+\beta-\kappaR-1$  in decreasing order.

The next lemma follows by iteratively applying the arguments we used to derive~\eqref{eq:RRR4} and \eqref{eq:r4}. 

 \bigskip
\begin{lemma}\label{Lem:Group4}
Source messages $M_{\i+\beta-\kappaR-1}^{\textnormal{Rx}}, M_{\i+\beta-\kappaR},\ldots, M_{\i+\beta-2}$ are successfully decoded  with probability tending to 1 as $n\to \infty$, whenever
 \begin{equation*}
 R_k < \frac{1}{2} \log\left(1+ \alpha_{k+1}^2 P\right), \qquad \forall k \in \{\i+\beta-\kappaR,\ldots, { \i+\beta -2}\}, \end{equation*}   
 \begin{equation*}
  R_{\i+\beta-\kappaR-1}^{\textnormal{Rx}} < \frac{1}{2} \log\left(1+ \alpha_{\i+\beta-\kappaR}^2 P\right),
  \end{equation*}
  and 
  \begin{equation*}
  R_{\Rx}>\max\Big\{  R_{\i+\beta-\kappaR-1}^{\textnormal{Rx}} ,\ R_{\i+\beta-\kappaR},\  \ldots,\ R_{\i+\beta -2}\Big\}.
  \end{equation*}
  \end{lemma}

\subsection{Transmission of source messages $M_{\i+\beta-\kappa_\Rx-\kappaT}, \ldots, M_{\i+\beta-\kappa_\Rx-\kappaT-2}, M_{\i+\beta-\kappa_\Rx-1}^{\textnormal{Tx}}$ (messages in group 3)}\label{sec:Group3}

Fix $k=\i+\beta-\kappaR-1$, and consider the special Tx/Rx pair $k$. (In Figure~\ref{Fig:Subnet1} this is the blue transmitter.) As we described in the previous subsection, Tx~$k$ already encoded its source message $M_{k}^{\textnormal{Rx}}$  into its input signal $X_{k}^n$ that was drawn from an iid Gaussian codebook.

Tx~$k$ now encodes its second source message~$M_{k}^{\textnormal{Tx}}$ which it transmits over the path
\begin{equation*}
\textnormal{Tx}~k \quad \longrightarrow \textnormal{Tx}~k-1 \quad \longrightarrow \textnormal{Rx}~k.
\end{equation*}
For convenience we will denote $M_{\i+\beta-\kappaR-1}^{\textnormal{Tx}}$ simply by $M_k$. Rx~$k$ will construct a  transmit signal $\Phi_k^n(M_k)$ and send a rate-$\frac{1}{2}\log(1+P)$ quantisation of ${\Phi}_k^n$ over the conferencing link to its left-neighbour Tx~$k-1$. This latter will then reconstruct the quantised sequence $\hat{\Phi}_{k}^n$ and send it over the network, see \eqref{eq:inputskminus} ahead.

We first describe the quantisation, and then the construction of $\Phi_k^n(M_k)$. Since $\Phi_k^n(M_k)$ won't be iid Gaussian, we will draw our quantisation codebook uniform over a sphere. 
Construct a random quantisation codebook 
\begin{equation*}
\s{C}_{\text{RD}, k} 
:= \Big\{ 
\h{\Phi}_{k}^n(u) 
\Big\}_{u = 1}^{\lfloor 2^{n R_{\Tx}}\rfloor}
\end{equation*}
by choosing all vectors iid uniformly over an $n$-dimensional sphere of radius $\sqrt{n\alpha_k^2 P}$.
 Tx~$k$  looks through $\s{C}_{\text{RD}, k}$ for the quantisation vector $\hat{\Phi}^n_{k}(u^*)$ whos angle with $\Phi^n_{k}$ is closest to $\sqrt{\frac{P}{P+1}}$:
 \begin{equation}
 u^* =\textnormal{argmin} \bigg| \angle\big(\h{\Phi}_{k}^n(u),{\Phi}_{k}^n\big)- \sqrt{\frac{P}{P+1}}\bigg|,
 \end{equation} 
 where the argmin is over all $u\in\big\{1,\ldots, 2^{n R_\Tx}\big\}$.
  Tx~$k$ sends index $u^*$ to its left-neighbour Tx~$k$ in Tx-conferencing round~1:
 \begin{equation}
 U\bs{1}_{k \to k-1} = u^*.
 \end{equation}
 
  By standard arguments (see e.g., proof of (134) in \cite{lapidothtinguelyMAC}), whenever 
 \begin{equation}\label{eq:rr1133}
  R_\Tx> \frac{1}{2}\log(1+P),
 \end{equation}
then
 \begin{equation}\label{eq:co}
 \bigg| \angle(\h{\Phi}_{k}^n(u^*),{\Phi}_{k}^n)- \sqrt{\frac{P}{P+1}}\bigg| \to 0 \quad \textnormal{in probability}.
 \end{equation} 

 
  After obtaining conferencing message  $ U\bs{1}_{k \to k-1}$, Tx~$k-1$ reconstructs   $\hat{\Phi}_k\big(U\bs{1}_{k \to k-1} \big)$ and transmits 
  \begin{equation}\label{eq:inputskminus}
  X_{k-1}^n = \alpha_{k}^{-1} \hat{\Phi}_k\big(U\bs{1}_{k \to k-1} \big)
  \end{equation}
  over the interference network.
  
  Rx~$k$ observes the channel outputs
 \begin{IEEEeqnarray}{rCl}\label{eq:outputskk}
Y_k^n &=& X_k^n+ \alpha_{k} X_{k-1}^n +Z_{k}^n \nonumber \\ &= &X_{k}^n+\h{\Phi}^n_{k} + Z_{k}^n 
\end{IEEEeqnarray}
which we choose to write
 \begin{IEEEeqnarray}{rCl}\label{eq:outputskk55}
Y_k^n &= &\underbrace{ \frac{P}{P+1}\Phi_k^n(M_k)}_{\textnormal{desired signal}}+
\underbrace{X_{k}^n}_{\textnormal{state sequence $S_k^n$}}+\underbrace{\Big(\h{\Phi}^n_k-\frac{P}{P+1}\Phi_k^n\Big)+Z_{k}^n}_{\textnormal{noise sequence $\tilde{Z}_k^n$}}.
\end{IEEEeqnarray}

Transmitter~$k$ encodes $M_k$ using a generalized dirty-paper code over spheres \cite{cohenlapidoth02-2,brosslapidothwigger12} of power $\alpha_k^2 \frac{P^2}{P+1}$ for the state-dependent channel in \eqref{eq:outputskk55}, i.e., for a channel with additive noise 
\begin{equation}
\tilde{Z}_k^n:= \Big(\h{\Phi}^n_k-\frac{P}{P+1}\Phi_k^n\Big)+Z_{k}^n,
\end{equation}
 and iid additive Gaussian state 
 \begin{equation}
 S_k^n:=X_k^n.
 \end{equation}
Let $\Psi_{\textnormal{DPC},k}^n(M_k)$ denote the resulting dirty-paper sequence. Tx~$k$ sets
\begin{equation*}
\Phi_k(M_k) := \frac{P+1}{P} \Psi_k^n(M_k).
\end{equation*}
Receiver~$k$ observes 
 \begin{IEEEeqnarray}{rCl}\label{eq:outputskk55}
Y_k^n &= &\underbrace{\Psi_{\textnormal{DPC},k}^n(M_k)}_{\textnormal{dirty-paper signal}}+
\underbrace{X_{k}^n}_{\textnormal{state sequence $S_k^n$}}+\underbrace{\Big(\h{\Phi}^n_k-\frac{P}{P+1}\Phi_k^n\Big)+Z_{k}^n}_{\textnormal{noise sequence $\tilde{Z}_k^n$}},
\end{IEEEeqnarray} 
and applies dirty-paper decoding over spheres to decode Message $M_k$.
 From a slight extension of Corollary~IV.2   
\cite{brosslapidothwigger12}\footnote{The extension is required because the state sequence $S_k^n$ is not uniform over a sphere and because the noise is not independent of the input. We however have the following limits (in probability): $\frac{1}{n}\|\Phi_k^n(M_k)\| \ \longrightarrow \  \alpha_k^2 (P+1)$ and $\frac{1}{n}\|S_k^n\|^2 \ \longrightarrow \ P$. Moreover when \eqref{eq:rr1133} holds, then by \eqref{eq:co} also 
$\frac{1}{n}\|\tilde{Z}_k^n\|^2 \ \longrightarrow \ 1+\alpha_k^2 \frac{P}{P+1}$, $\frac{1}{n} < \tilde{Z}_k^n, S_k^n> \ \longrightarrow \ 0$, and $\frac{1}{n} < \tilde{Z}_k^n, \Psi_{\textnormal{DPC},k}^n> \ \longrightarrow \  0$ (all in probability). Under these assumptions, Corrollary~IV.2 in \cite{brosslapidothwigger12} can be shown to extend readily, e.g., using the arguments in the proof of Remark~III-5 in \cite{brosslapidothwigger12}.}
%
%
we obtain that, if \eqref{eq:rr1133} holds and if
  \begin{equation}\label{eq:Rate3}
 R_k^{\textnormal{Tx}} <\frac{1}{2} \log\left( 1+ \frac{\alpha_k^2 \frac{P^2}{P+1}}{1+\alpha_k^2\frac{P}{P+1}}\right),
  \end{equation}  
  then $\Pr[\h{M}_{k} \neq M_{k}] \to 0$ as $n\to \infty$.\\

 The same process is repeated for  $k= \i+\beta-\kappaR-2, \ldots, \i+\beta-\kappaR-\kappaT$  in decreasing order. \\
 


 The next lemma follows by iteratively applying the arguments we used to derive~\eqref{eq:Rate3}, see also \eqref{eq:rr1133}.

 \bigskip
\begin{lemma}\label{Lem:Group3}
Source messages $M_{\i+\beta-\kappaR-\kappaT}\ldots, M_{\i+\beta-\kappaR-2}, M_{\i+\beta-\kappaR-1}^{\Tx}$ are successfully decoded  with probability tending to 1 as $n\to \infty$, whenever
\begin{IEEEeqnarray*}{rCl}
R_k < \frac{1}{2} \log\left( 1+ \frac{\alpha_k^2 \frac{P^2}{P+1}}{1+\alpha_k^2\frac{P}{P+1}}\right), \hspace{1cm} k\in\{\i+\beta-\kappaR-\kappaT,\ldots,\i+\beta-\kappaR-2\}\IEEEeqnarraynumspace 
\end{IEEEeqnarray*}
and 
\begin{equation*}
R_{\i+\beta-\kappaR-1}^{\Tx} < \frac{1}{2} \log\left( 1+ \frac{ \frac{P^2}{P+1} \alpha_{\i+\beta-\kappaR-1}^2}{1+ \frac{P}{P+1}\alpha_{\i+\beta-\kappaR-1}^2}\right).
 \end{equation*}
  and 
  \begin{equation*}
  R_{\Tx}>\frac{1}{2}\log(1+P).
  \end{equation*}
\end{lemma}

\bibliographystyle{ieeetr}

\begin{thebibliography}{20}

\bibitem{willems83}
F.~M.~J. Willems, ``The discrete memoryless multiple access channel with partially cooperating encoders,''
\emph{IEEE Trans. Inform. Theory}, vol.~29, no.~3, pp.~441--445, Nov. 1983.

\bibitem{slepianwolf73} D. Slepian and J.K. Wolf, ``A coding theorem
  for multiple access channels with correlated sources,'' \emph{Bell
  System Tech. J.} vol. 52, pp. 1037--1076, Sep., 1973.

  \bibitem{brosslapidothwigger08}
S. Bross, A. Lapidoth, and M. A. Wigger,   ``The Gaussian MAC
 with conferencing encoders,'' in \emph{proc.~Intl.~Symp.~Inform.~Theory (ISIT),} pp. 2702--2706, Toronto, Canada, July 6--11,
 2008. 
 
    \bibitem{brosslapidothwigger12} S. I. Bross, A. Lapidoth and M. Wigger, ``Dirty-paper coding for the Gaussian multiaccess
channel with conferencing,"  \emph{IEEE Trans.~Inform.~Theory}, vol.~58, no.~9, pp.~5640--5668, Sep.,~2012. 

\bibitem{maricyateskramer07}
I.~Maric, R.~D. Yates, and G.~Kramer, ``Capacity of interference channels with partial transmitter cooperation,'' \emph{IEEE Trans.~Inform.~Theory}, vol.~53, no.~10, Oct., 2007.

\bibitem{cohenlapidoth02-2}
Aaron~S. Cohen and Amos Lapidoth.
\newblock The {Gaussian} watermarking game.
\newblock {\em IEEE Trans. Inform. Theory}, 48(6):1639--1667, June 2002.

\bibitem{costa83}
Max H.~M. Costa.
\newblock Writing on dirty paper.
\newblock {\em IEEE Trans. Inform. Theory}, 29(3):439--441, May 1983.



\bibitem{simeoneetal08} O.~Simeone, O.~Somekh, G.~Kramer, H.~V.~Poor,
  and S~Shamai (Shitz), ``Three-User Gaussian Multiple Access Channel
  with Partially Cooperating Encoders,'' in \emph{proc.  Asilomar
  Conference Signals, Systems and Computers}, Oct. 2008.  

\bibitem{kramerwigger09}
M. A. Wigger and G. Kramer, ``Three-User MIMO MACs with
  Cooperation,'' in \emph{Proc.~Inform.~Theory Workshop (ITW)}, pp.~221--225, Volos, Greece,
  Jun., 2009. 

\bibitem{daboraservetto06-1} R. Dabora and S. Servetto, ``A multi-step conference for cooperative broadcast," in \emph{Proc. Intl.~Symp.~Inform. Theory (ISIT)}, pp.~2190--2194, Seattle, WA, Jul., 2006.

\bibitem{daboraservetto06-2} R. Dabora and S. Servetto, ``Broadcast channels with cooperating decoders," \emph{IEEE Trans.~Inform.~Theory,} vol. 52, no. 12, pp. 5438--5454, Dec. 2006.


 \bibitem{ihsiang-tx} I.-H. Wang and D. N. C. Tse, ``Interference mitigation through limited transmitter cooperation," \emph{IEEE Trans. on Inf. Theory,} vol.~57, no. 5, pp.~2941--2965, May 2011.
 
 \bibitem{ihsiang-rx} I.-H. Wang and D. N. C. Tse, "Interference mitigation through limited receiver cooperation,"  \emph{IEEE Trans. on Inf. Theory,} vol. 57, no. 5, pp. 2913--2940, May 2011. 

\bibitem{yossi} Y. Steinberg, ``Instances of the relay-broadcast channel and cooperation strategies," in \emph{Proc.  Intl.~Symp.~Inform. Theory (ISIT)},
pp.~2653--2657, Hong Kong, June 2015.

\bibitem{mao} H. Mao, W. Feng, and N. Ge, ``Receiver cooperation for MIMO broadcast channels with finite-rate feedback," \emph{IEEE Communications Letters}, pp.~887--890, vol~19, no.~5, May~2015. 

\bibitem{sadaf} S. Salehkalaibar and M. R. Aref, ``On the capacity region of a class of Z channels with cooperation," in \emph{Proc. 2010 Intl. Symp. Inform. Theory and its Appl. (ISITA)}, Taichung, China, 17--20 Oct. 2010, pp.~464--468.

\bibitem{ziv} Z. Goldfeld, H. H. Permuter, G. Kramer, ``Duality of a source coding problem and the semi-deterministic broadcast channel with rate-limited cooperation." Online: \texttt{http://arxiv.org/abs/1405.7812}


\bibitem{ng} C. T. K. Ng, A. J. G. I. Maric, and R. D. Y. S. Shamai, ``Iterative and
one-shot conferencing in relay channels" in \emph{Proc. of Inform. Theory Workshop 2006}, Punta del Este, Uruguay, 13--17 March 2006, pp.~193--197. 

\bibitem{maricyateskramer07} I. Maric, R. D. Yates, and G. Kramer, ``Capacity of interference channels with partial transmitter cooperation,"  \emph{IEEE Trans.~Inform.~Theory,} vol. 53, no. 10, Oct. 2007.

\bibitem{levyshamai09} N. Levy and S. Shamai (Shitz), ``Clustered local decoding for Wyner-
type cellular models," \emph{IEEE Trans.~Inform.~Theory,} 
vol. 55, no. 11, Nov. 2009, pp. 4976--4985.

\bibitem{lapidothshamaiwigger07} A. Lapidoth, S. Shamai (Shitz), and M. Wigger, ``On cognitive interference networks," in \emph{Proc. of Inform. Theory Workshop (ITW)},  Lake Tahoe, USA, Sep. 2--7, 2007.

\bibitem{ntranosmaddah-alicaire14-1}
%
V.~Ntranos, M.~A.~Maddah-Ali, and G.~Caire,
``Cellular interference alignment,"
 \emph{IEEE Trans.~Inform.~Theory,} vol. 61, no. 3, March 2015,
pp.~1194--1217.

\bibitem{ntranosmaddah-alicaire15-2}
V. Ntranos, M. A. Maddah-Ali, and G. Caire, 
``Omni-directional antennas and
asymmetric configurations," \emph{IEEE Trans.~Inform.~Theory,} vol. 61, no. 12, March 2015, pp.~6663 - 6679.

\bibitem{ntranosmaddah-alicaire15-3}
V. Ntranos, M. A. Maddah-Ali, and G. Caire, 
``On uplink-downlink duality for cellular IA,"July 2014.
Online: \texttt{http://arxiv.org/abs/1407.3538}.

\bibitem{biao} B. He, N. Yang, X. Zhou, and J. Yuan,
``Base station cooperation for confidential
broadcasting in multi-cell networks,"
\emph{IEEE Trans. on Wireless Communications,} vol.~14, no.~10, Oct.~2015, 
pp. 5287--5299.


\bibitem{bandeelgamalveeravalli15} M.~Bande, A.~El Gamal, and V.~V.~Veeravalli, 
``Flexible backhaul design with cooperative transmission in cellular
interference networks,"  in \emph{Proc. 2015 Intl. Symp. Inform. Theory}, Hong Kong,  China, June~14--19, 2015.

\bibitem{cadambejafar08}
S.~A.~Jafar and V.~R.~Cadambe, ``Interference alignment and degrees of freedom of
  the k -user interference channel,'' \emph{IEEE Trans. Inform. Theory}, vol.~54, pp.~3425--3441, Aug. 2008.


\bibitem{Wyner-94}
A.~D. Wyner, ``Shannon-theoretic approach to a {G}aussian cellular
  multiple-access channel,'' {\em IEEE Trans.~Inform.~Theory},
  vol.~40, pp.~1713--1727, Nov.,~1994.

\bibitem{Hanly-Whiting-93}
S.~V. Hanly and P.~A. Whiting, ``Information-theoretic capacity of
  multi-receiver networks,'' {\em Telecommunication Systems}, vol.~1,
  pp.~1--42, 1993.

\bibitem{rate-limited}
S.~{Shamai (Shitz)} and M. Wigger, ``Rate-limited transmitter-cooperation in Wyner's asymmetric interference network ," in {\em proc. Intl.~Symp.~Inform.~Theory (ISIT)}, pp.~425--429, St. Petersburg, Russia, Jul.~31 -- Aug.~5, 2011.
  
  
\bibitem{lapidothlevyshamaiwigger14} A.~Lapidoth, N.~Levy, S.~Shamai (Shitz), and M.~Wigger, ``Cognitive Wyner networks with clustered decoding,"  \emph{IEEE Trans. Inform. Theory}, vol.~60, no.~10, pp.~6342-6367 Oct. 2014. 

\bibitem{Simeone-2011-Tut}
O.~Simeone, N.~Levy, A.~Sanderovich, O.~Somekh, B.~M.~Zaidel, H.~V.~Poor and S.~Shamai (Shitz),
``Cooperative Wireless Cellular Systems: An Information-Theoretic View,'' \emph{Foundations and Trends in Communications and Information Theory (FnT),} Vol. 8, No.~1-2, 2011, pp. 1--177, Now Publishers, 2012.

\bibitem{coverthomas06}
 T. Cover and J. Thomas, ``Elements of information theory," 2nd edition, \emph{Wiley-Interscience}.


\bibitem{elgamalkim10} A. El Gamal and Y.-H. Kim, ``Network  information theory," \emph{Cambridge}.

\bibitem{lapidothtinguelyMAC}
A.~Lapidoth and S.~Tinguely, ``Sending a bivariate {G}aussian over a {G}aussian
  {MAC},'' \emph{{IEEE} Transactions on Information Theory}, vol.~56, no.~6, 
  pp. 2714--2752, Jun. 2010.


 \end{thebibliography}

\end{document}